\documentclass[accepted]{melba}

\usepackage{mwe} 

\usepackage{amsmath,amsfonts}

\usepackage{subfigure} 
\usepackage{float}
\usepackage{booktabs}
\usepackage{multirow}
\usepackage{multicol}
\usepackage{bbding}
\usepackage{makecell}
\usepackage{diagbox}

\usepackage{pifont}
\usepackage{rotating}

\newcommand{\Target}{\mathrm{Target}}

\newcommand{\ratio}{\mathrm{ratio}}

\newcommand{\Stolen}{\mathrm{Stolen}}
\newcommand{\zlatent}{Z_{\mathrm{latent}}}
\newcommand{\zhyperlatent}{Z_{\mathrm{hyper}}}

\newcommand{\DD}{\mathrm{D}}
\newcommand{\D}{\mathrm{D1}}
\newcommand{\PP}{\mathrm{P}}

\melbaid{2025:032}  
\doi{10.59275/j.melba.2025-113f}
\melbaauthors{Li, Ayache and Delingette}  
\email{huiyu.li@inria.fr}
\volume{3}
\firstpageno{728}  
\melbayear{2025}  
\datesubmitted{2024-08-30}  
\datepublished{2025-12-05}  

\melbaspecialissue{Medical Imaging with Deep Learning (MIDL) 2020}
\melbaspecialissueeditors{Marleen de Bruijne, Tal Arbel, Ismail Ben Ayed, Hervé Lombaert}

\ShortHeadings{Data Exfiltration by Compression Attack}{Li, Ayache and Delingette}

\title{Data Exfiltration by Compression Attack: Definition and Evaluation on Medical Image Data}

\author{
	\firstname Huiyu \surname Li\aff{1},
	\name Nicholas \surname Ayache\aff{1},
    \name Hervé \surname Delingette\aff{1}
}
\affiliations{
	\num 1 \addr Inria, Epione Team, Université Côte d’Azur, Sophia Antipolis, France
}

\abstract{
With the rapid expansion of data lakes storing health data and hosting AI algorithms, a prominent concern arises: how safe is it to export machine learning models from these data lakes? In particular, deep network models, widely used for health data processing, encode information from their training dataset, potentially leading to the leakage of sensitive information upon its export. This paper thoroughly examines this issue in the context of medical imaging data and introduces a novel data exfiltration attack based on image compression techniques.
This attack, termed {\em{Data Exfiltration by Compression}}, requires only  access to a data lake and is based on lossless or lossy image compression methods. 
Unlike previous data exfiltration attacks, it is compatible with any image processing task and depends solely on an exported network model without requiring any additional information to be collected during the training process. We explore various scenarios, and techniques to limit the size of the exported model and conceal the compression codes within the network.
Using two public datasets of CT and MR images, we demonstrate that this attack can effectively steal medical images and reconstruct them outside the data lake with high fidelity, achieving an optimal balance between compression and reconstruction quality. Additionally, we investigate the impact of basic differential privacy measures, such as adding Gaussian noise to the model parameters, to prevent the Data Exfiltration by Compression Attack. We also show how the attacker can make their attack resilient to differential privacy at the expense of decreasing the number of stolen images. Lastly, we propose an alternative prevention strategy by fine-tuning the model to be exported.
 }


\begin{document}

\twocolumn[\maketitle]

\section{Introduction}
The increasing use of AI technology on medical data for clinical research has stimulated the establishment of medical data warehouses or data lakes, where structured or unstructured data produced by major hospitals and health organizations, are stored and analyzed. Of course, the access to these data lakes is heavily restricted and regulated, typically only allowing a remote access  to external users. In general, remote data scientists are given access to a secured space where data curation and the training of machine learning algorithms can be performed. 
 
The leakage of privacy-sensitive medical data from these data spaces poses a serious threat to the reputation of the health organizations managing such a data lake. Cybercriminals could exploit data leaks to harm third parties or ransom the health organization. To protect the privacy and security of these data lakes, it is crucial for data owners to understand their potential vulnerabilities, prevent privacy attacks~\cite{kaviani2022adversarial, zhang2022deep} and to develop effective mitigation measures.

Privacy attacks can be categorized into two main groups based on their objectives, as shown in Table~\ref{tab1}. All attacks consider a \emph{target dataset}, which is the dataset that the attacker wants to steal or compromise, and a \emph{target model}, which is the machine learning model that the attacker aims to compromise or exploit.
The first group of attacks focuses on the recovery of specific properties of the target dataset. Specifically, 
\textbf{Property Inference Attacks}~\cite{zhikun2022inference, wang2022poisoning, wang2022group, zhang2021leakage, melis2019exploiting, ganju2018property, ateniese2015hacking} aim to infer additional properties of the target dataset, i.e., that are not explicitly correlated to the learning task, using only the parameters of the target model as prior knowledge. 
For instance, this includes the inference of  the gender distribution in a patient dataset when the learning task is tumor segmentation.
The \textbf{Attribute Reconstruction Attacks} (aka feature re-derivation, attribute inference)~\cite{chen2022practical, kaissis2020secure} aim to infer sensitive attributes of a target data sample from the knowledge of the target model, its output, and other non-sensitive attributes. 

In a \textbf{Membership Inference Attacks} (aka tracing attack)~\cite{10285881, 10214061, 10262058, hu2022membership} , the attacker attempts to infer whether a particular data sample was used in the target dataset to train the target model. 
\textbf{Re-identification Attacks}~\cite{chen2023data, 9432915, el2011systematic} determine an individual’s identity by linking the information from an pseudo-anonymised personal data to another dataset in which the same individual is contained.

\begin{table*}[htbp]
\caption{Privacy attacks categorized based on their target objectives: (1) targeting the recovery of specific properties (rows 1-4) and (2) targeting the entire target data (rows 5-7). These attacks depend on the attacker's knowledge about the target model or the output of the model on the target data.}\label{tab1}
\fontsize{8.8pt}{8.8pt}\selectfont
\centering
\begin{tabular}{lccccc}
\toprule
\multirow{3}*{Attack} &\multicolumn{4}{c}{Adversary Knowledge} &{Adversary Output}\\
\cmidrule(r){2-5}
&\multicolumn{2}{c}{Target Model} &{Target} &\multirow{2}*{Additional}\\
\cmidrule(r){2-3}
& \emph{Architecture} &\emph{Parameters} &{Output}\\
\midrule
\makecell[l]{Property Inference Attacks \\ \cite{zhikun2022inference}} &\XSolid &\Checkmark &\XSolid &\XSolid &\makecell[c]{Properties of \\ target data}\\
\makecell[l]{Attribute Reconstruction Attacks\\ \cite{chen2022practical}} &\Checkmark &\Checkmark &\Checkmark &\XSolid &{Individual features}\\
\makecell[l]{Membership Inference Attacks \\ \cite{10285881}} &\Checkmark &\Checkmark &\Checkmark &\XSolid &\makecell[c]{Membership of \\ target data}\\
\makecell[l]{Re-identification Attacks \\ \cite{chen2023data}} &\XSolid &\XSolid &\XSolid &\XSolid &{Individual identity}\\
\midrule
\makecell[l]{Model Inversion Attacks \\ \cite{10003239}} &\Checkmark &\Checkmark &\Checkmark &\XSolid &\makecell[c]{Reconstructed \\ target data}\\
\makecell[l]{Gradient Inversion Attacks \\ \cite{10209197}} &\Checkmark &\XSolid &\XSolid &Gradients &\makecell[c]{Reconstructed \\ target data}\\
\makecell[l]{\textbf{Data Exfiltration}\\ \cite{li2022data}} &\XSolid &\XSolid &\XSolid &\makecell[c]{Compression \\ Codes} &\makecell[c]{Reconstructed \\ target data}\\
\bottomrule
\end{tabular}
\end{table*}

Our work is primarily related to the second group of attacks, which aim at recovering the entire target dataset. These attacks exploit the output data generated by machine learning models to estimate the input  target data. More precisely, 
\textbf{Gradient Inversion Attacks}~(GIA)~\cite{10209197, hatamizadeh2023gradient, hatamizadeh2022gradvit, huang2021evaluating} leverage the gradients of the loss function with respect to the weights of the target model in order to reconstruct the training data. They  exploit the  gradient information  to understand the relationship between the model's output and  input data, ultimately reconstructing the input data after benefiting from  prior knowledge of the data domain.
Different from GIA, \textbf{Model Inversion Attacks} (MIA)~\cite{10003239, nguyen2024label, chen2021knowledge, fredrikson2015model, 10184490}, aka \textbf{Attribute Inference Attacks}, aim to reconstruct the target data from the target model's output. Algorithmically, MIAs are designed as solving an optimization problem by estimating  the input data that maximizes the likelihood of the output data. 

\textbf{Data Exfiltration Attacks}~\cite{ULLAH201818} (DEA) encompass various methods to leak data outside  a data warehouse by an insider or an outsider of the data management organization. When dealing with machine learning algorithms, data exfiltration attacks aim at recovering the training data following the export of the model. Since  a neural network, e.g. an autoencoder, somewhat memorizes its training data within its weights and as a result it is  expected that the export of that  network can be maliciously utilized  to steal data. For instance, the technique of steganography was used  to hide data or even malware~\cite{StegoNet} inside the least significant bits of a neural network. Also, model inversion attacks solely based on the network weights was proposed in~\cite{NEURIPS2022_90692737}. Under certain hypotheses about the network architecture (lack of skip connections) and training method (gradient based), the authors show that it is possible to reconstruct the training data given the trained model. However, the quality of the reconstructed training data is fairly limited and the proposed method is not suitable for U-net type image segmentation networks that include skip connections. Deep generative image models such as GANs or diffusion models are also likely to release their training data by using specific membership inference attacks~\cite{10.5555/3620237.3620531}. The authors show that GANs are less likely to generate their training data than diffusion models and that only a subset (typically 20\%) of the original training data can be recovered. While the quality of recovered training images can be high, generative image models are by essence likely to lead to data theft and are therefore unlikely to be exported outside any data lake. Finally, transpose attack~\cite{amit2024transposeattackstealingdatasets}, was recently introduced to recover training image data by running a classification network backward. This approach was successfully tested of small size images and avoids using multiple heads as in multi-task learning to hide a malicious model. 

The  GIA, MIA, and DEA  aim at reconstructing the input training data, but they suffer from a number of  shortcomings that limit their effectiveness in the context of medical data lakes. 
First, GIA and MIA require the attacker to have access to the gradient or the output of the target model but without accessing the input / target data. This situation is fairly uncommon except during the training stage of a network using federated learning \cite{li2020review}. Second, the quality of the reconstructed images is intrinsically limited for model inversion attacks or degrades rapidly with the number of memorized images increases for the transpose attack. Third, the attacks require that network to solve a  specific task. For instance, model inversion and transpose attacks require the export of  classification networks. Image generative models can exfiltrate high quality images but do not solve classical image processing tasks and can be easily detected. Finally, none of the previous attacks have been tested or evaluated on medical image datasets, which are typically large, often three-dimensional, and highly sensitive.  

In this paper,  we introduce a novel attack coined as ``\textbf{Data Exfiltration by Compression Attack}'' which is widely applicable within medical data lakes, and that can lead to catastrophic data leakage. The Data Exfiltration by Compression (DEC) attack is a data exfiltration attack which is based on  learned deep image compression networks (Fig.~\ref{fig1}). The attacker compresses the target data into  codes and embeds them within the exported neural network, enabling the reconstruction of the target dataset outside the data lake. Despite the straightforward principle, the DEC attacker must cope with several  contradictory objectives:  efficiently solving  an image processing task, maximizing the number of stolen images, minimizing the size of the exported network, hiding the compression codes inside the network, and ensuring  the compression codes are resistant to noise in the model. 

In the DEC attack, as with most other data exfiltration attacks,  the exported network is intentionally designed to leak data whereas data leakage occurs unintentionally for the  MIA and GIA.
But unlike existing data exfiltration attacks, it can generate high quality and high resolution volumetric images and it is  independent from the network architecture, and the task achieved by the model.  Similarly to the transpose attack, in the external pre-training scenario, the exported network architecture is single headed and therefore cannot be discovered easily by the data owner. Besides, it is also resilient to the addition of Gaussian noise up to a certain level.  

The DEC attack is evaluated in this paper in the context of medical image analysis with the performance of image segmentation tasks. Please note that stealing medical images from data lakes is fairly challenging due to the size of the datasets (several tens of megabytes) and their subtle content. But the DEC attack principles could be extended to electronic health records, biosignals (e.g. ECG data), or biological data.

In this paper, we have studied the different constraints associated with the DEC attack and proposed several mitigation plans. Our contributions can be summarized as follows:

\begin{enumerate}
    \item We introduce a novel attack based on data compression which is agnostic to the architecture and task of the exported network. We introduce two distinct scenarios where the DEC attack is applicable, carefully balancing their inherent pros and cons. We explore how various hypotheses can influence the design of the attack. These hypotheses  include the compression methodology (lossy vs. lossless), the ability to import a pretrained model into the data lake, the method of inserting  compression codes into the model, the addition of noise to the model, and the application of common mitigation measures.
     
     \item We demonstrate that learned image compression methods,  such as those in~\cite{47602}, are well suited for: i)  generating high-quality medical images with small compression codes, and ii) creating  compression codes that are resilient to the addition of Gaussian noise in the network. 
     
    \item We  conducted extensive   experiments in real-world scenarios on two public datasets (LiTS and BraTS), focusing on stealing CT or MR images embedded within a segmentation network. To the best of our knowledge, this is the first time an exfiltration attack has been evaluated on a medical image dataset. We monitored the number and  quality of stolen images, the efficacy of the segmentation network, and the size of the exported network. Our results show  that the DEC attack poses a significant threat to medical data lakes, enabling  attackers to efficiently balance a high number of stolen images with strong network performance in solving image segmentation tasks.

\end{enumerate}

This work extends our conference paper ~\cite{li2022data} in several significant ways. First, we introduce the External Pre-training Scenario (Section~\ref{EPscenario}) to avoid the necessity of exporting attack model from the data lake. Second, we improve attack efficiency by substantially reducing the size of both the compression codes and the exported model (Section~\ref{sizeReduction}). Third, we incorporate steganography as a new approach to conceal compression codes within exported models. Finally, we introduce a set of mitigation strategies (Section~\ref{mitigation}) aimed at limiting the risk of data exfiltration, and we analyze the robustness of the DEC attack under differential privacy (Section~\ref{DPmitigation}).

\section{Method}
Data Exfiltration by Compression attack leads to the unauthorized leakage  of sensitive  data from a remote data lake. The attacker who carries out the attack,  has access to the data lake, but with malicious intentions. The harmful objectives of the attacker may be for instance to steal data in order to create an external high-value dataset, to damage the data owner's reputation by publishing the stolen data, or to ransom the data owner. The attacker has access to the data lake either after getting an authorization by the data owner or after stealing the identity of an honest data lake user. Each data lake user has access to a secured demilitarized zone  where data curation, model learning and testing take place without any possibility of exporting data. When a machine learning model reaches a certain performance level, the user can ask the data owner to export the trained model in order to test it on other proprietary or public data. The data owner must then decide what action should be taken to limit the risk of data leakage outside of the data lake following the export of the trained model. This paper aims at evaluating the risk of data leakage of medical images from a healthcare organization and at providing mitigation actions for the data owner. 

\begin{figure}[htbp]
\centering
\includegraphics[width=1.\linewidth]{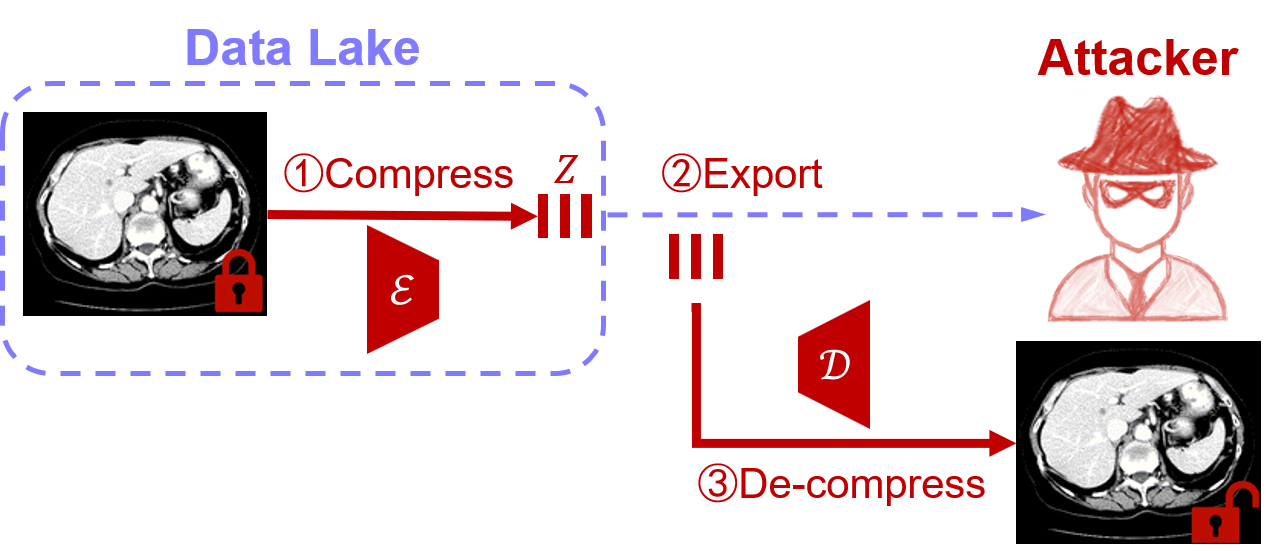}
\caption{Overview of Data Exfiltration by Compression Attack.}
\label{fig1}
\end{figure}

The attack is based on data compression and takes place in three stages when exporting a trained model as seen in Fig.~\ref{fig1}.
First, the attacker encodes the target data into compression codes $Z$, and then  the compression codes and the  network are exported outside the data lake. Finally, the attacker uses a decoder to decompress the compression codes, reconstructing the data into its original resolution and format.

Furthermore, to be realistic,  the exported machine learning model should have  a significant performance in order to convince the data owner that the original objective for accessing the data lake has been reached. The exported model solves a  \emph{utility task} which can be for instance an  image segmentation, registration, detection or classification algorithm when processing medical images. A main advantage of the proposed attack is that the success of the attack does not depend on the chosen task or the network architecture.

We first present in Section~\ref{scenarios} the two main scenarios of the attack  depending whether  the attacker is able to import a compression encoder or not.
We then detail several key aspects such as the lossy compression network (in Section~\ref{lossyCompression}), the reduction of the exported model size (in Section~\ref{sizeReduction}), the hiding of compression codes and network models (in Section~\ref{hidingCodesModels}), the mitigation strategies (in Section~\ref{mitigation}), and the method to produce noise resilient attacks (in Section~\ref{DPmitigation}).

\subsection{Data Exfiltration by Compression Attack: Two Main Scenarios}\label{scenarios}

\begin{figure*}[htbp]
\centering
\includegraphics[width=1.\linewidth]{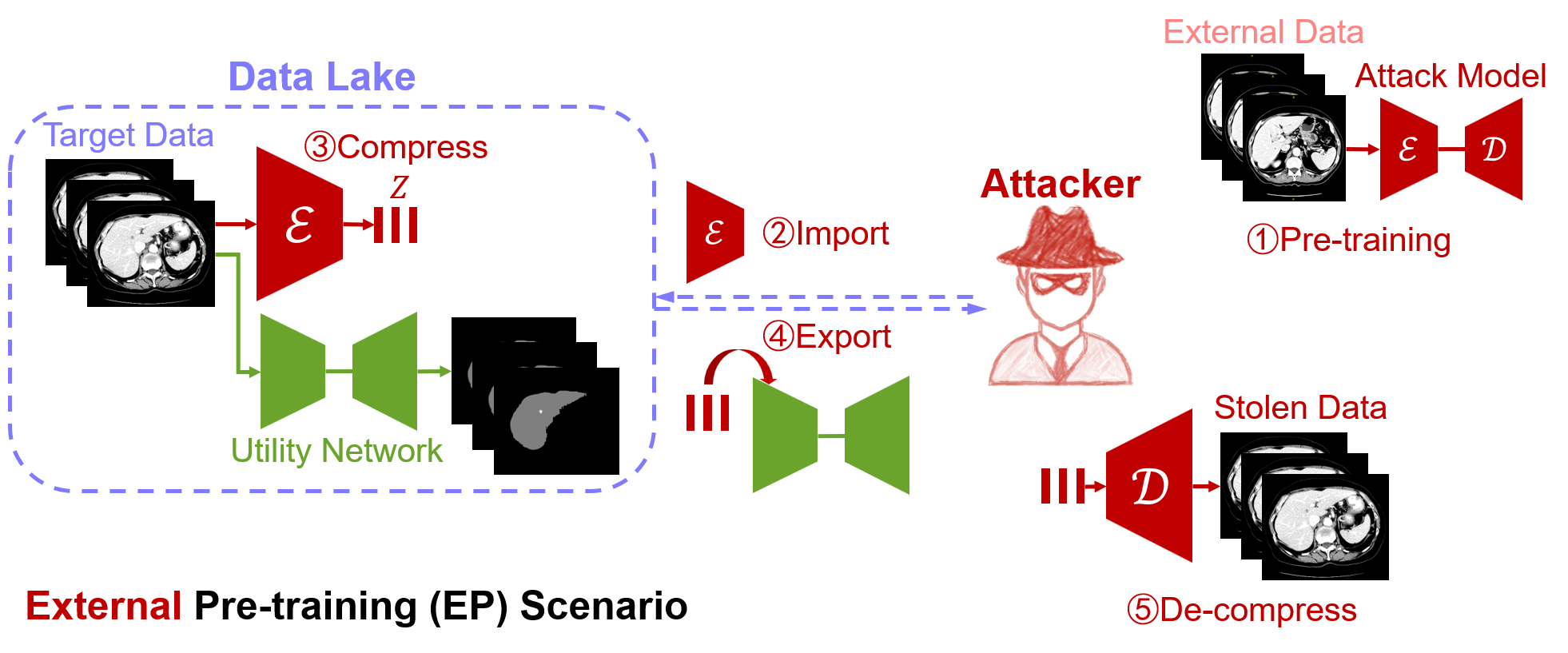}
    \caption{The pipeline of lossy compression based attack in External   Pre-training Scenario, where the data owner has access to an encoder-decoder pair outside the data lake. Initially, an encoder-decoder pair is trained on an external dataset. The trained encoder is then imported into the data lake. Next, the encoder compresses the target data into compression codes $Z$ while a utility network is simultaneously trained to conceal the attack behavior. Subsequently, the attacker exports the compressed codes embedded within the trained utility network from the data lake. Finally, with the exported $Z$ and the trained decoder outside the data lake, the attacker can de-compress the stolen dataset.
} 
\label{fig2}
\end{figure*}
\begin{figure*}[htbp]
\centering
\includegraphics[width=1.\linewidth]{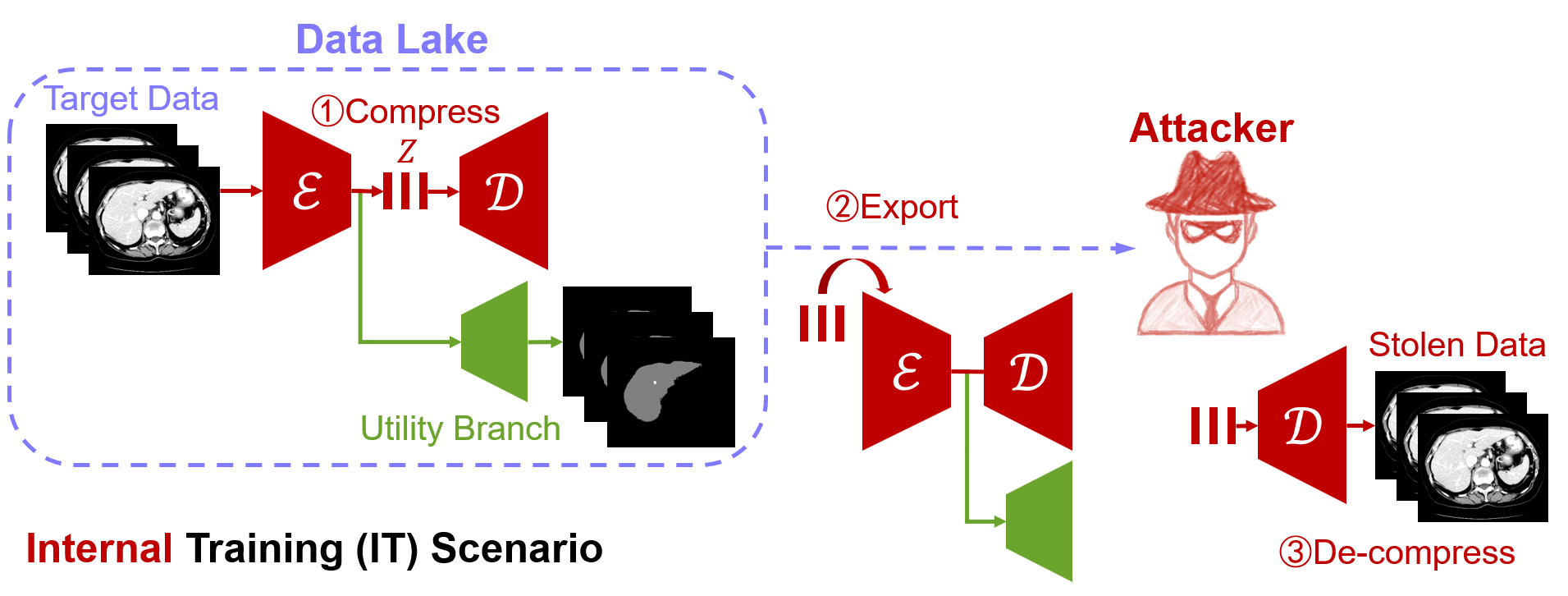}
\caption{The pipeline of lossy compression based attack in Internal Training Scenario, without access to an external encoder-decoder pair. First, an attack model trained inside the data lake compresses the target data into compression codes $Z$, utilizing a shared encoder with a utility decoder branch for hiding the attack behavior. Then, the attacker exports the utility network, attack decoder, and $Z$ from the data lake. Finally, the stolen data can be de-compressed outside the data lake using the exported Z and the decoder.}
\label{fig3}
\end{figure*}

\textbf{External Pre-training (EP) Scenario.} \label{EPscenario}
In this first scenario, depicted in Fig.~\ref{fig2}, the attacker creates an encoder-decoder pair outside the data lake, eventually by training the pair on an external dataset. 
Then, the data owner allows the attacker to import the encoder into the data lake, and the target data is then compressed by the encoder. 
To export the compression codes,  the attacker trains a utility network that  reaches the necessary performance and uses the network to hide the compression codes (see section~\ref{hidingCodesModels} for more details) before the export.
The attacker can then use the decoder outside the data lake to reconstruct the target data.

Thus, in this scenario, the attacker does not need to train and to export the decoder network outside the data lake. On the downside, the quality of the data reconstruction may suffer from a potential domain shift between the target  and  the external dataset from which the encoder was trained. Another issue for the attacker is to hide from the data owner, the real nature of the encoder network. The attacker may argue for importing existing backbone or foundation models to boost the model performance or can resort to more sophisticated techniques such as hiding a model inside another model~\cite{guo2021hidingneuralnetworksinside}.

Lossless compression-based attack represents a specific configuration within this EP Scenario. In such case, the encoder-decoder pair can be a standard compression tool, such as ZIP / UNZIP. Note however that lossless image compression methods have inherent limitations in terms of compression ratio which restricts the amount of data that can be efficiently compressed. Nonetheless, in the remainder,  we use the lossless compression-based attack as a baseline for comparison with lossy compression approaches.

\textbf{Internal Training (IT) Scenario.}
In this second scenario (Fig.~\ref{fig3}), we consider that the attacker does not have any access to a pretrained encoder-decoder pair, unlike in the previous scenario. Instead, the attacker  trains the encoder-decoder pair inside the data lake, potentially utilizing adversarial autoencoders like VAE-GAN~\cite{8957359}, deep compression networks~\cite{agustsson2019generative}, or a combination of both~\cite{mentzer2020high}.  The attacker can then use the locally trained encoder to compress the target data. At this stage, the attacker must achieve three objectives: i) train a convincing utility model solving a specific task, ii) export the compressed data with the decoder, and iii) export the utility model. 
We propose an approach based on multi-task learning which solves the 3 objectives as can be seen in Fig.~\ref{fig3}. More precisely,  the utility model consists of the trained encoder as the backbone model combined with a dedicated \emph{utility branch} to solve the utility task. The second branch, including the decoder, has the unique purpose of reconstructing the target data outside the data lake. An alternative design would be to create a utility network distinct from the encoder-decoder pair but this would have the disadvantage of potentially increasing significantly the size of the exported model. In~\cite{amit2024transposeattackstealingdatasets}, the authors argue that multi-headed networks can be easily detected as suspicious by the data owner. However, there are various ways to hide the network structure as discussed in~\cite{guo2021hidingneuralnetworksinside} for instance.  
After training the utility branch, the attacker can hide the compression codes inside the two-headed network and ask for the export of the resulting network to the data owner. Outside the data lake, the attacker can reconstruct the target data by leveraging the decoder branch and the compression codes.
 
This internal training scenario has the advantage of training a data-specific encoder-decoder pair that is not subject to any domain shift. Please note that in this scenario, creating models that overfit the training data is a desirable property. However, this scenario requires the export of the decoder network which increases the risk of theft detection due to the increased size of the network and the two-heads in the network architecture. In Section \ref{sizeReduction}, we explore various approaches to reduce the size of the decoder and the compression codes.

\subsection{Lossy Compression  and Utility Models}\label{lossyCompression}
We detail below the lossy image compression model and utility branch / model implemented  in this paper.
Motivated by the success of learned image compression techniques~\cite{agustsson2019generative}, we  use  the High Fidelity Compression (HiFiC) model proposed in~\cite{mentzer2020high} which combines GAN with learning based compression. The model includes an encoder that transforms an image $x$ into a latent code $y=E(x)$, and a decoder or generator that converts the latent code $y$ into an approximation of the target image, $x'=D(y)\approx x$. Relying on adversarial training,  a discriminator $\hat{D}(x')$ is used to estimate if the generated image is real or fake. 

Thus the resulting architecture based on an encoder $E$, decoder $D$ (aka the generator) and discriminator $\hat{D}$, is in some ways a mixture of a GAN (with $D$ and $\hat{D}$) with an auto-encoder (with $E$ and $D$). 
This architecture is suitable for learned image compression with high quality data reconstruction and it differs from other attack models~\cite{10003239} (such as MIA) that are based on a GAN model with random noise as input. 

\begin{figure}[htbp]
\centering
\includegraphics[width=1.\linewidth]{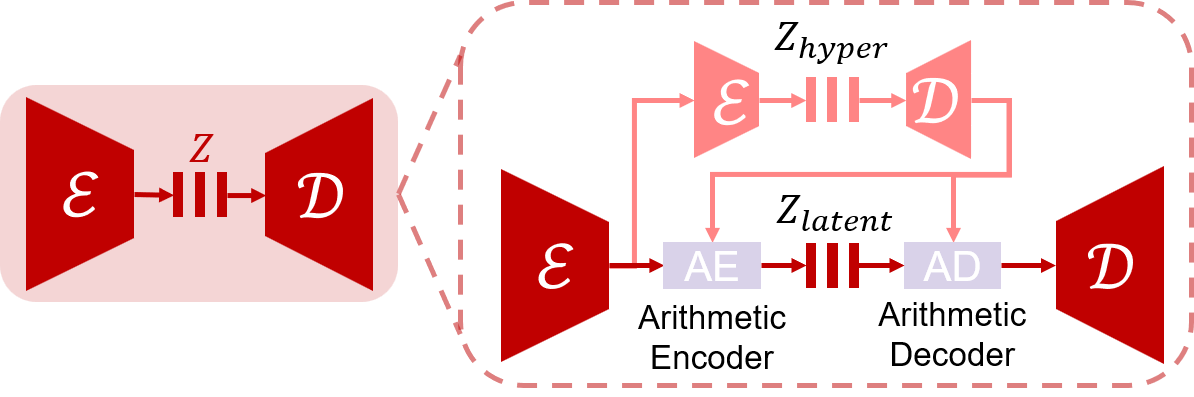}
\caption{Operational diagrams of the learned image compression method using an hyperprior entropy model \cite{47602}.
The upper section corresponds to an hyperprior auto-encoder while the  lower one shows an image auto-encoder
architecture  where AE, AD represent respectively arithmetic encoder and arithmetic decoder.
}
\label{fig4}
\end{figure}

The HiFiC model uses a hyperprior entropy model~\cite{47602}, which helps to predict the distribution of the latent code (see 
Fig.~\ref{fig4}). In fact, the (decoded) latent code $y$ is generated by an arithmetic decoder $AD$, which takes as input the latent variable $\zlatent$ and is controlled by the transformed  hyperlatent variables $\zhyperlatent$. As a result it is sufficient to store the pair of latent and hyperlatent variables $Z=[\zlatent,\zhyperlatent]$ in order to generate the code $y$ which then generates an image.

\subsection{Size Reduction  of Compression Codes  and Decoder Model}\label{sizeReduction}

The core function of the attack model in the DEC attack is to compress original medical images into a compact representation and subsequently reconstruct them with high fidelity. Importantly, the proposed DEC attack is network-agnostic and can employ any model that performs this transformation and reconstruction process, such as autoencoders or latent diffusion models. The choice of architecture is guided by the need to balance compression efficiency—maximizing the number of images that can be stored within limited capacity—with reconstruction fidelity, ensuring that the reconstructed images remain diagnostically useful.

For the utility task, we primarily focus on image segmentation  unlike most existing attacks that require solving classification tasks. Regarding the utility network in EP Scenario (Fig.~\ref{fig2}), we employ a lightweight U-Net with 4 levels as the utility network, which takes target images as input and produces segmentation masks. In the IT Scenario (Fig.~\ref{fig3}),  the attacker  uses the encoder of the generative compression model as the feature extraction network and trains a dedicated decoding branch specifically designed for the segmentation  task. In this case the utility branch is the decoder part of a lightweight U-Net  without any skip connections. This branch takes as input the output of the encoder, before the arithmetic encoding.

To limit the risk of theft detection by the data owner, it is important  to minimize the size of the compression codes in both scenarios, and the size of  the decoder  in the IT Scenario. To reduce the size of the compression codes $Z$, we  can decrease the number of  channels in both the hyperlatent $\zhyperlatent$ and latent $\zlatent$. This reduction of the dimension of the latent code directly decreases the overall size of the compression codes but also may potentially impact the quality of the reconstructed images. 

To reduce the size of the decoder network $D$, we explored various classical model pruning techniques, including Structured pruning~\cite{anwar2017structured}, Unstructured pruning~\cite{cheng2017survey}, Quantization~\cite{yang2019quantization}, and Knowledge Distillation (KD)~\cite{gou2021knowledge}. However, Structured pruning, Unstructured pruning, and Quantization failed to create  effective networks due to the complexity of our model compared to the ones  typically used for parameter pruning. Similarly, KD did not produce any satisfactory results, potentially due to the fact that we were distilling a generative model instead of a classification model, as is commonly done.

Given the limitations of network pruning in our case, we opted for a more direct strategy. The original decoder $D$ in the HiFiC model ~\cite{mentzer2020high}, is quite large with a size of 627 MB. A careful analysis of the architecture of D led to the observation that its size is mainly determined by nine repeating ResNet blocks. We propose to replace the original decoder $D$ with a  smaller one $D1$, where the nine ResNet blocks have been removed, resulting in a much reduced model size of 29 MB. 

\subsection{Hiding Compression Codes in Exported Models}\label{hidingCodesModels}
The data exfiltration by compression attack assumes that compression codes are hidden inside exported network weight files. More precisely, depending on the size of exported network, the compression codes can be hidden either in the least significant bits of the parameters or as an extra key-value pair in the parameter dictionary.

Indeed, if  the attack model is large enough, it is possible to apply  steganography~\cite{ingemar2008digital} by storing some chunks of the codes in  the least significant bits of the network weight file.
After exporting the network outside the data lake, the compression codes can be easily extracted from the least significant bits of the checkpoint files, allowing the decoder to reconstruct the images. Using  steganography has two  major advantages. First, it does not affect  the exported network size, and second it makes it very difficult for the data owner to detect the hidden codes.

When the exported model is small, steganography does not allow one to hide enough data. In such case, the compression codes are stored in the checkpoint file  as additional entries in a dictionary with dedicated keys, assuming, for instance, an HDF5 file format.  
This approach has the drawback of increasing the size of the exported network due to the added compression codes which may raise suspicion from the data owner if the total size becomes excessive. Also, it is possible that the added dictionary entries may be detected since they are not connected to other nodes or layers in the exported model.

\subsection{Mitigation Measures}\label{mitigation}
To limit the risk of data exfiltration by compression attack or to detect the attack, data owners can apply a number of  measures listed below.

\textbf{DEC Attack Prevention.} An obvious way to prevent the DEC attack is to forbid the export of any machine learning model outside the data lake. This drastic decision, however, is often not acceptable for the remote user who has developed the model.  Alternatively, the data owner can implement several actions that limit the risk of DEC attack without eliminating it. First,  the data owner should verify the integrity and honesty of the users that are granted access to the data lake. In addition, multi-factor authentication, adding an extra layer of security by requiring multiple forms of verification, should limit the risk of an attacker to steal the  identity of an honest user. Third, data owners should  closely monitor the machine learning models that are imported inside the data lake. Indeed the external pre-training scenario shows that importing a data compression encoder in the data lake makes the DEC attack very efficient (no decoder to export) and  difficult to detect. Data owners should verify the origin of the model to be imported and its nature. Fourth, the data owner may control the last iterations of the model training (fine-tuning)  before exporting the binary files outside the data lake. This allows the owner to verify that all weights of the model have been modified during training and that no additional data has been maliciously inserted into the model. This has the advantage of being relatively simple to implement by the data owner as it  can be largely automated. A drawback is that the remote user has no control over the performance of the exported network.

Finally, the data owner can apply  Differential Privacy (DP)~\cite {dwork2006differential}  by systematically adding noise to the exported model. The added noise aims to degrade or destroy the compression codes potentially stored in the exported model. Indeed,  the addition of noise to lossless compression files makes the decompression impossible. Similarly, the noise added to the compression codes $Z$ is also expected to have a significant effect since $\zlatent$ is the input of an arithmetic encoder which is very noise sensitive.  It is important to note however, that the application of DP must be calibrated, since it effectively enhances privacy at the cost of  degrading the performance of the exported  utility model. In Section~\ref{DPmitigation}, we discuss how the attacker can adapt its methodology to cope with the addition of noise in the exported model. 

\textbf{DEC Attack Detection.}  The data owner can implement actions in order to detect that a remote user is performing a DEC attack. A first method consists in checking the source code to find any  manipulation of the neural networks such as the application of steganography or the addition of data in the dictionary file. This is obviously difficult to implement due to the time and expertise required. The second way to detect the attack is to filter the exported networks depending on their size, and to carefully check the content of large neural networks that are exiting the data lake. Indeed, small networks cannot embed data of large size such as medical images.

\subsection{DEC Attack Resisting to Differential Privacy} \label{DPmitigation}

Knowing that the data owner is applying DP to the exported model, the attacker might opt for a compression method that is resilient to the addition of white noise, even if it means increasing the code size. 

As discussed previously, the compression codes $Z = [\zhyperlatent, \zlatent]$ are sensitive to  noise addition since it can disrupt the precise intervals used for arithmetic decoding, potentially leading to  the impossibility to produce any decoded data. 
However, the decoded latent code $y$ following the arithmetic decoder, as seen in Fig.~\ref{fig4}, is by design much less  sensitive to noise since it was shown in~\cite{47602} that $y$ distribution closely resembles a Gaussian distribution. As a result, an effective strategy for the attacker to make network model resilient to Gaussian noise is to encode each image using the decoded latent variable $y$ instead of $Z$, although its size is much larger than that of $Z$. In this case, the attacker sacrifices the quantity of the stolen images in favor of increasing their quality.

\section{Implementation Details}\label{sec3}
\subsection{Dataset}
To evaluate the feasibility of the proposed attack, we focus on a use case involving the storage of CT or MRI images in a medical data lake, with image segmentation as the designated utility task. Abdominal CT images are fairly large volumetric images (typically $512\times 512\times 100$) encoded in 2 bytes per voxel and therefore are particularly  challenging to exfiltrate. 

The MICCAI 2017 Liver Tumor Segmentation (LiTS) Challenge dataset~\cite{bilic2023liver} contains 130 abdominal CT cases for training and 70 CTs for testing. In this dataset, the utility task is to segment the liver in a supervised manner. The Multimodal Brain Tumor Segmentation (BraTS) Challenge 2021  dataset~\cite{baid2021rsna} includes 1251 skull stripped brain MR sequences for training and 219 cases for validation with the segmentation of the whole tumor from  FLAIR MR sequences as the utility task.
On the LiTS (resp. BraTS) dataset, we randomly partition the training set into 104/13/13 (resp. 1000/126/125) images used for training, validation, and testing of the utility task. Also, for testing the attack model, we use the 70 (resp. 219) test images in the LiTS (resp. BraTS) dataset.

In the EP Scenario, two external datasets are required to pre-train the attack model outside the data lake which should closely resemble the two target dataset in order to minimize domain shift between the target and external datasets. 
As the external CT dataset, we chose the 2021 Fast and Low-resource semi-supervised Abdominal oRgan sEgmentation (FLARE)~\cite{ma2022fast} dataset containing 361 abdominal CT cases for training and 50 abdominal CTs for testing. Similarly for MR, we used the 2022 Brain Tumor Sequence Registration (BraTS-Reg) Challenge~\cite{baheti2021brain}, having 120 cases for training, 20 cases for validation and 20 cases for testing. 
We used the same training and validation / testing sets as those in the original FLARE and BraTS-Reg datasets.

\subsubsection{Pre and Post-processing}
In the LiTS and FLARE datasets, the network input size is $512\times512$ pixels which is the same as the slice resolution. In the BraTS and BraTS-Reg dataset, edge padding is applied since the slice resolution is only $240\times240$ pixels. Finally, a min-max intensity normalization is applied on the whole image. Since the HiFiC compression model takes RGB images as input,  each slice is surrounded by its upper and lower slices to fill the three channels. For post-processing and display, all decompressed  images are mapped back to their original minimum and maximum range.

\begin{figure*}[htbp]
	\centering
	\subfigure[CT images.]{
	\includegraphics[width=0.48\linewidth]{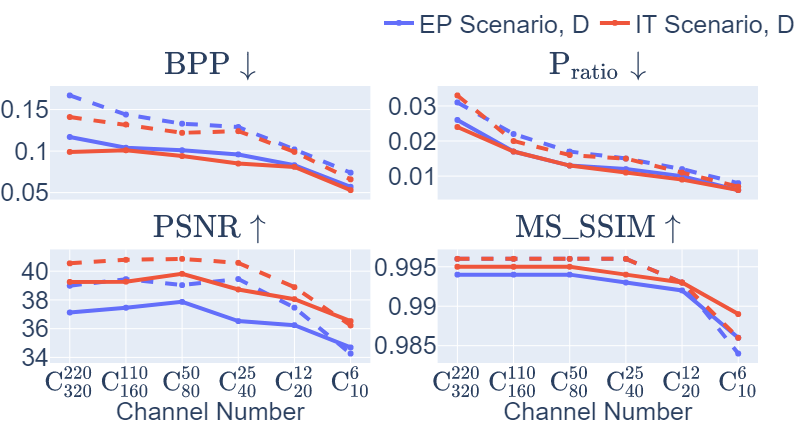}
	}\label{fig5a}
	\subfigure[MRI images.]{
    	\includegraphics[width=0.48\linewidth]{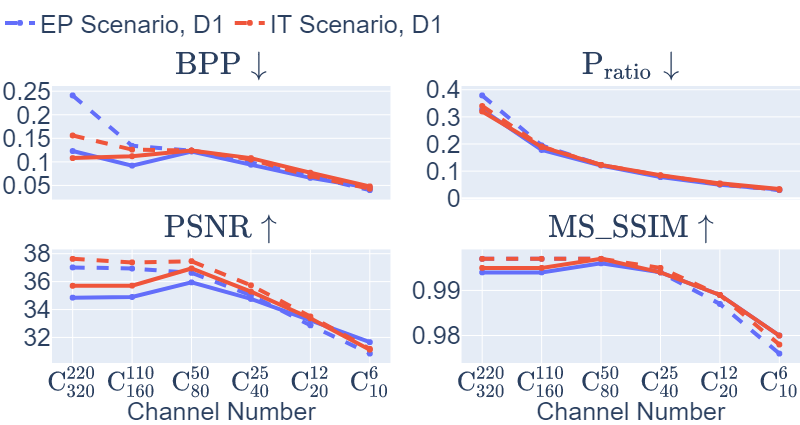}
	}\label{fig5b}
	\caption{Lossy compression based attack on CT images (a) and MRI images (b) with various channel numbers $C^{|\zlatent|}_{|\zhyperlatent|}$ in x axis. The compression efficiency is measured using the criteria of BPP and $\PP_{\ratio}$ in the first row, while the reconstruction fidelity is assessed using the criteria of PSNR and MS\_SSIM in the second row.}
	\label{fig5}
\end{figure*}

\subsection{Evaluation Metrics}
To assess compression efficiency of the attack model, we used metrics such as BPP (bits per pixel) and a self-defined $\PP_{\ratio}$ which  represents the ratio between the disk size of the compression codes generated by lossy compression and the disk size of the baseline lossless compression (as given by \emph{gzip 3.6.0}). The metrics  $\PP_{\ratio}$  provides  an indication of the effectiveness of lossy compression compared to lossless compression.
Additionally, to assess the reconstruction fidelity of the decompressed images compared to the original ones, we make use of the  PSNR (Peak Signal-to-Noise Ratio) and MS\_SSIM (Multi-Scale Structural Similarity)~\cite{wang2003multiscale} metrics.
Finally, we evaluate the image segmentation performance of the utility model with the Dice score.

\subsection{Compression and utility Models}
The HiFiC image compression  model is borrowed from~\cite{mentzer2020high} while  the utility model is a simplified version of the  network in \cite{li2020deep}.  The same HiFiC model architecture is used in both the external pre-training and  internal training  scenarios.   
All  models are optimized with the Adam~\cite{kingma2014adam} algorithm and training continues untill validation loss has converged.

\section{Results}\label{sec4}
\subsection{Attack Effectiveness}\label{sec4.2}

We assess the potential trade-offs faced by the attacker when crafting their attack model. These trade-offs encompass maximizing the quality of the compressed images, minimizing the size of the compression codes, minimizing the overall model size, and ensuring the utility task is effectively fulfilled.

To optimize those metrics, the attacker can modify the architecture of the encoder-decoder model, and select a specific exfiltration scenario.
Regarding the model architecture, the attacker may either choose the full size original decoder $D$ (627 MB) or its reduced version $D1$ (29 MB) (see Section~\ref{sizeReduction}). In addition, the number of latent and hyperlatent channels may be decreased to reduce the compression code size. We write $C^{|\zlatent|}_{|\zhyperlatent|}$ as the network configuration where the number of latent (resp. hyperlatent) channels is $|\zlatent|$ (resp. $|\zhyperlatent|$).

In Fig.~\ref{fig5},  the attack performance is evaluated in terms of compression efficiency (BPP, $\PP_{\ratio}$) and reconstruction fidelity (PSNR, SSIM) across different network architectures and scenarios. The key results are summarized below.

\textbf{EP Scenario vs. IT Scenario.} 
The internal training (IT) scenario leads to improved image quality compared to the external pre-training (EP) scenario. This can be explained by the potential domain shift between the external dataset and the target dataset. In terms of model size, however, the two scenarios are equivalent.

\textbf{Type of Decoder: D vs D1.}
In Fig.~\ref{fig5}, we observe that the reduced decoder D1 (29 MB) surprisingly outperforms the original large decoder D (627 MB) both in terms of quality of reconstruction and compression efficiency, and also for  both scenarios,
Thus, stripping the ResNet blocks from decoder D leads to improved image quality, which can be explained by a better balance in terms of model size  among the encoder, decoder, and discriminator components.

\textbf{Latent Channel Number Configurations.} 
As the number of channels of the latent/hyperlatent variables decreases, we observe an increase in the compression efficiency of the attack model but also  a reduction of the reconstruction fidelity. The quality of reconstructed images decreases more sharply when the channel number is equal to or below  $\mathrm{C^{25}_{40}}$. 
Thus, we found that setting the channel numbers at $\mathrm{C^{50}_{80}}$ resulted in a good trade-off   between compression efficiency and reconstruction fidelity. 
input

\begin{figure}[htbp]
	\centering
	\subfigure[CT images.]{
		\includegraphics[width=\linewidth]{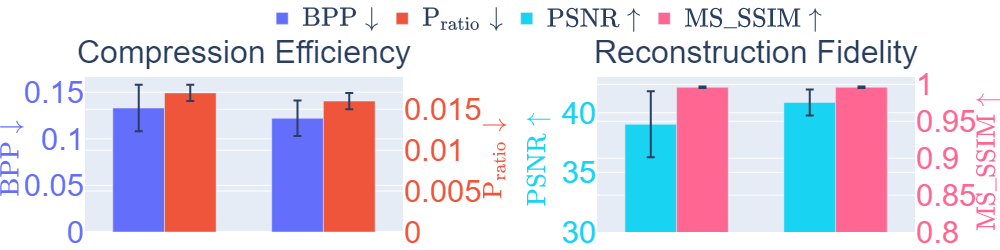} 
	}\label{fig6a}
	\subfigure[MRI images.]{
    	\includegraphics[width=\linewidth]{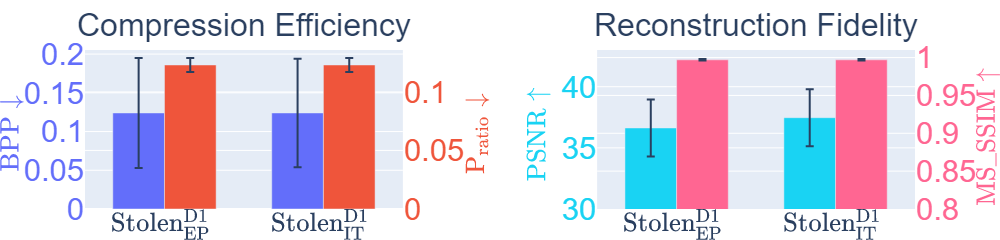}
	}\label{fig6b}
	\caption{Lossy compression based attack on CT images (a) and MRI images (b) with a specific channel numbers ($\mathrm{C^{50}_{80}}$) for the EP and IT Scenarios. $\Stolen^{\D}_{\mathrm{EP}}$ denotes the decompressed images in the EP Scenario with a reduced D1 decoder. }
	\label{fig6}
\end{figure}

\textbf{Compression-Fidelity Compromise.} 
Based on the previous results, we have selected the decoder $D1$ and the configuration of the latent and hyperlatent variables $\mathrm{C^{50}_{80}}$, as the optimal architecture of the HiFiC encoder-decoder pair. 
In Fig.~\ref{fig6}, we display  more specifically  the compression and reconstruction quality performances. In terms of reconstruction quality, we obtain a PSNR of approximately 40 for CT images and around 38 for MRI images while  the MS\_SSIM values are  close to 1. This indicates an  excellent perceptual quality of the reconstructed images that are hardly discernible from the original ones. In terms of compression efficiency,  the $\PP_{\ratio}$ for CT images is approximately 0.015, indicating that the lossy image compression-based attack generated compressed images are 67 times smaller than those produced by the lossless zipped image compression-based attack. For MRI images,  the $\PP_{\ratio}$ is  around 0.12,  10 times higher than that of CT images, which  can be attributed to the presence of a large uniform background in the skull-stripped original MR images.

\begin{figure}[htbp]
	\centering
	\subfigure{
	\rotatebox{90}{\scriptsize{~~~~~~~~~~CT}}
		\begin{minipage}[t]{0.17\linewidth}
			\centering
			\includegraphics[width=1\linewidth]{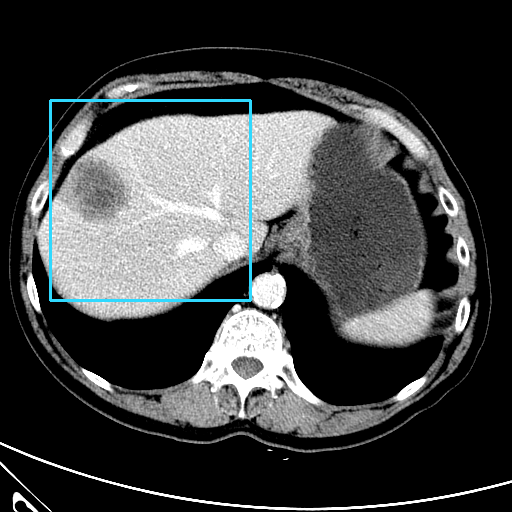}
		\end{minipage}
	}\hspace{-0.5\baselineskip}\vspace{-0.5\baselineskip}
	\subfigure{
		\begin{minipage}[t]{0.17\linewidth}
			\centering
			\includegraphics[width=1\linewidth]{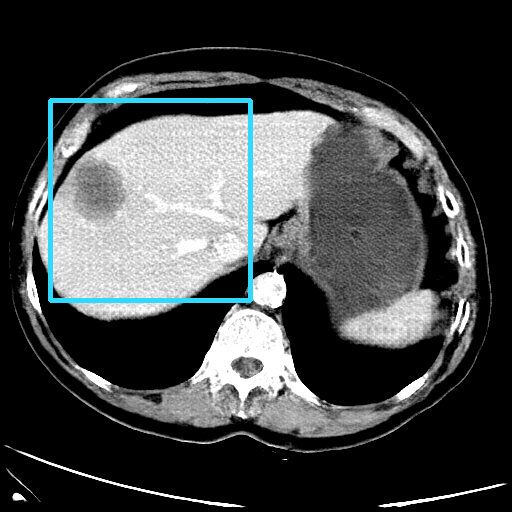}
		\end{minipage}
	}\hspace{-0.5\baselineskip}
	\subfigure{
		\begin{minipage}[t]{0.17\linewidth}
			\centering
			\includegraphics[width=1\linewidth]{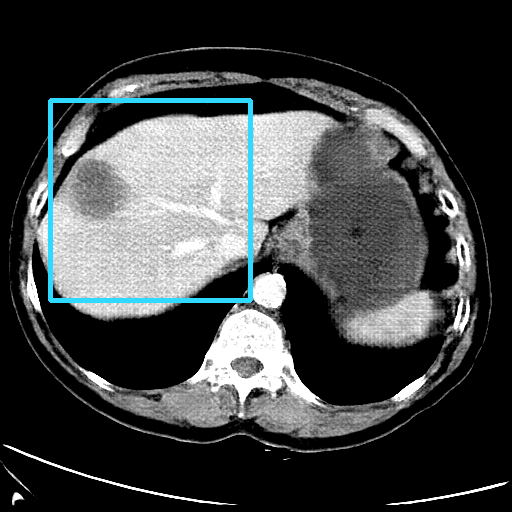}
		\end{minipage}
	}\hspace{-0.5\baselineskip}
	\subfigure{
		\begin{minipage}[t]{0.17\linewidth}
			\centering
			\includegraphics[width=1\linewidth]{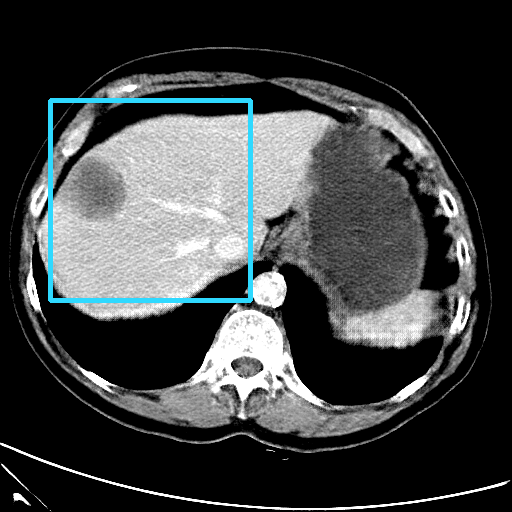}
		\end{minipage}
	}\hspace{-0.5\baselineskip}
	\subfigure{
		\begin{minipage}[t]{0.17\linewidth}
			\centering
			\includegraphics[width=1\linewidth]{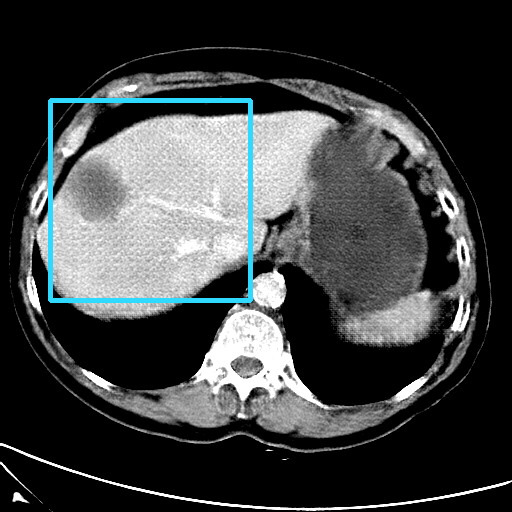}
		\end{minipage}
	}
	
	\subfigure{
	\rotatebox{90}{\scriptsize{~~~~~~Zoomed}}
		\begin{minipage}[t]{0.17\linewidth}
			\centering
			\includegraphics[width=1\linewidth]{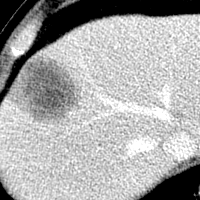}
		\end{minipage}
	}\hspace{-0.5\baselineskip}\vspace{-0.2\baselineskip}
	\subfigure{
		\begin{minipage}[t]{0.17\linewidth}
			\centering
			\includegraphics[width=1\linewidth]{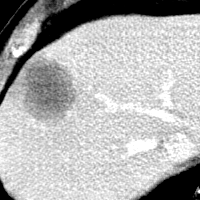}
		\end{minipage}
	}\hspace{-0.5\baselineskip}
	\subfigure{
		\begin{minipage}[t]{0.17\linewidth}
			\centering
			\includegraphics[width=1\linewidth]{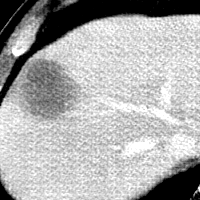}
		\end{minipage}
	}\hspace{-0.5\baselineskip}
	\subfigure{
		\begin{minipage}[t]{0.17\linewidth}
			\centering
			\includegraphics[width=1\linewidth]{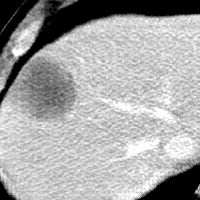}
		\end{minipage}
	}\hspace{-0.5\baselineskip}
	\subfigure{
		\begin{minipage}[t]{0.17\linewidth}
			\centering
			\includegraphics[width=1\linewidth]{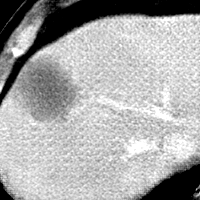}
		\end{minipage}
	}
    
	\subfigure{
	\rotatebox{90}{\scriptsize{~~~~~~~~MRI}}
		\begin{minipage}[t]{0.17\linewidth}
			\centering
			\includegraphics[width=1\linewidth]{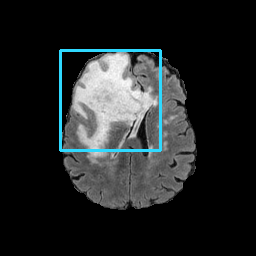}
		\end{minipage}
	}\hspace{-0.5\baselineskip}\vspace{-0.5\baselineskip}
	\subfigure{
		\begin{minipage}[t]{0.17\linewidth}
			\centering
			\includegraphics[width=1\linewidth]{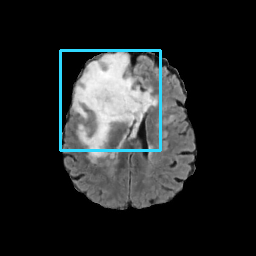}
		\end{minipage}
	}\hspace{-0.5\baselineskip}
	\subfigure{
		\begin{minipage}[t]{0.17\linewidth}
			\centering
			\includegraphics[width=1\linewidth]{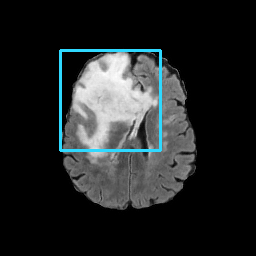}
		\end{minipage}
	}\hspace{-0.5\baselineskip}
	\subfigure{
		\begin{minipage}[t]{0.17\linewidth}
			\centering
			\includegraphics[width=1\linewidth]{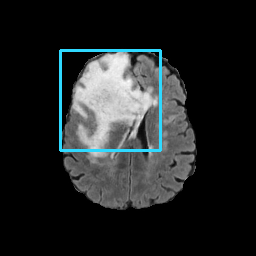}
		\end{minipage}
	}\hspace{-0.5\baselineskip}
	\subfigure{
		\begin{minipage}[t]{0.17\linewidth}
			\centering
			\includegraphics[width=1\linewidth]{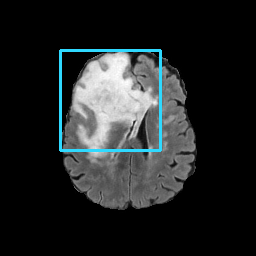}
		\end{minipage}
	}

	\subfigure{
	\rotatebox{90}{\scriptsize{~~~~~~Zoomed}}
		\begin{minipage}[t]{0.17\linewidth}
			\centering
			\includegraphics[width=1\linewidth]{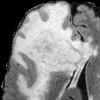}
			\centerline{\footnotesize $\Target$}
		\end{minipage}
	}\hspace{-0.5\baselineskip}
	\subfigure{
		\begin{minipage}[t]{0.17\linewidth}
			\centering
			\includegraphics[width=1\linewidth]{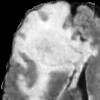}
			\centerline{\footnotesize ${\Stolen}^{\DD}_{\mathrm{EP}}$}
		\end{minipage}
	}\hspace{-0.5\baselineskip}
	\subfigure{
		\begin{minipage}[t]{0.17\linewidth}
			\centering
			\includegraphics[width=1\linewidth]{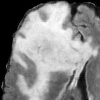}
			\centerline{\footnotesize ${\Stolen}^{\D}_{\mathrm{EP}}$}
		\end{minipage}
	}\hspace{-0.5\baselineskip}
	\subfigure{
		\begin{minipage}[t]{0.17\linewidth}
			\centering
			\includegraphics[width=1\linewidth]{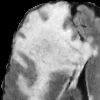}
			\centerline{\footnotesize ${\Stolen}^{\DD}_{\mathrm{IT}}$}
		\end{minipage}
	}\hspace{-0.5\baselineskip}
	\subfigure{
		\begin{minipage}[t]{0.17\linewidth}
			\centering
			\includegraphics[width=1\linewidth]{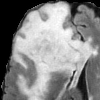}
			\centerline{\footnotesize ${\Stolen}^{\D}_{\mathrm{IT}}$}
		\end{minipage}
	}
	\caption{Lossy image reconstructions on CT (row 1, 2) and MRI (row 3, 4) images, where the row 2, 4 provide a zoomed-in view of the bounding box region of the row 1, 3. The leftmost column represents the target images, while the subsequent four columns show the stolen images reconstructed by the decoder D or D1 in two scenarios.
	}
	\label{fig7}
\end{figure}

A visual comparison between target and stolen images is available in Fig.~\ref{fig7}. 
We observe that the stolen images from IT Scenario closely resemble the input ones, particularly in the tumor regions, whereas stolen images from EP Scenario exhibit blurring artifacts in finer details. In both cases, the stolen images reconstructed by $D1$ demonstrate a comparable quality to those reconstructed by $D$, thus further confirming the effectiveness of the reduced decoder $D1$.

\subsection{Utility Task Performances}\label{sec4.3}
To solve the utility task, in EP Scenario, we employed a whole dedicated U-Net network (UN as Utility Network), whereas  in the IT Scenario, we utilized a Utility Branch (UB) network. These models were trained either on  the target dataset from the data lake or on the stolen dataset reconstructed by the attacker. Subsequently, they were tested on an unseen test dataset (a subset of the target dataset).
The UN model, trained on the target dataset, also serves as a baseline model for comparison. 

\begin{figure}[htbp]
	\centering
	\subfigure[CT images.]{
		\includegraphics[width=0.6\linewidth]{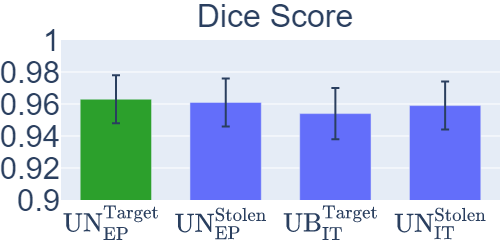}
	}\label{fig8a}
	\subfigure[MRI images.]{
    	\includegraphics[width=0.6\linewidth]{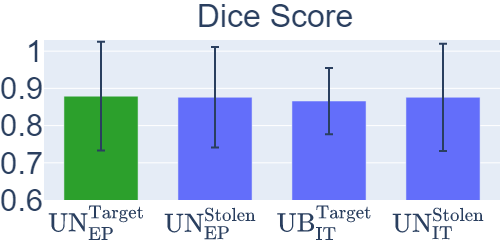}
	}\label{fig8b}
	\caption{Utility task results on CT and MRI images.
	The UN and the UB models trained on the target dataset ($\mathrm{UN^{Target}_{EP}}$, $\mathrm{UB^{Target}_{IT}}$) are utilized to to execute the utility task within the data lake.
	The UN model trained on the stolen dataset ($\mathrm{UN^{Stolen}_{EP}}$, $\mathrm{UN^{Stolen}_{IT}}$) is to evaluate the practical utility of the stolen dataset in solving the same utility task.
}
	\label{fig8}
\end{figure}

As shown in Fig.~\ref{fig8}, both the UN and UB models have good results  on both the target and stolen dataset. This means that in both scenarios the attacker can  provide  convincing evidences  to the data owner that the model to be exported is effective.
This also indicates that the stolen image data is indeed useful for training a model for solving the same utility task.

\subsection{Code Hiding Method}\label{sec4.4}
To hide the compression codes within the exported network, the attacker can either use steganography or add entries in  the file dictionary (see Section~\ref{hidingCodesModels}). 
The  choice may depend on the size of the decoder:  steganography is probably preferred if the decoder is large enough; otherwise, the codes are stored as additional dictionary entries. 

For the steganography method, the values of each parameter are converted from float32 into binary (32 bits), and we decided to allocate the least significant 16 bits to store the compression codes. 
This approach does not increase the size of the exported network, but it requires that the size of the exported network be at least twice as large as that of the compression codes. This is why it  is well suited in the case where  a  full-size decoder (D) is used for the IT Scenario. However, steganography modifies the parameter values of the model and it may impact the model accuracy if a large number of bits are allocated. It is therefore essential to verify  that the performance of the exported model is not significantly affected by steganography.

\begin{table*}[!htbp]
\caption{The size (MB) of encoded compression codes of decoder D in IT Scenario when stealing 100 images.}\label{tab2}
\centering
\begin{tabular}{l|c|c|c|c|c|c}
\hline
\multirow{2}*{Data} &\multicolumn{6}{c}{Model}\\
\cmidrule(r){2-7}
&$\mathrm{C^{220}_{320}}$ &$\mathrm{C^{110}_{160}}$ &$\mathrm{C^{50}_{80}}$ &$\mathrm{C^{25}_{40}}$ &$\mathrm{C^{12}_{20}}$ &$\mathrm{C^{6}_{10}}$\\
\hline
CT &230.504 &179.866 &153.824 &132.260 &121.610 &81.728\\
\hline
MRI &44.153 &28.425  &21.078 &15.771 &10.513 &7.318\\
\hline
\end{tabular}
\end{table*}

Table~\ref{tab2} shows that the size of the compression codes when stealing 100 images, is consistently smaller than half the size of D (627 MB), confirming that the full decoder $D$ has sufficient capacity for steganography.
Moreover, we found no noticeable degradation of the compression and reconstruction performance of the HiFiC  model when allocating 16 bits for data exfiltration. 
But when allocating  more than 16 bits,  image artifacts  start to appear in stolen images, confirming that this corresponds to a good compromise.

\begin{figure}[htbp]
\centering
	\subfigure[CT images.]{
		\includegraphics[width=\linewidth]{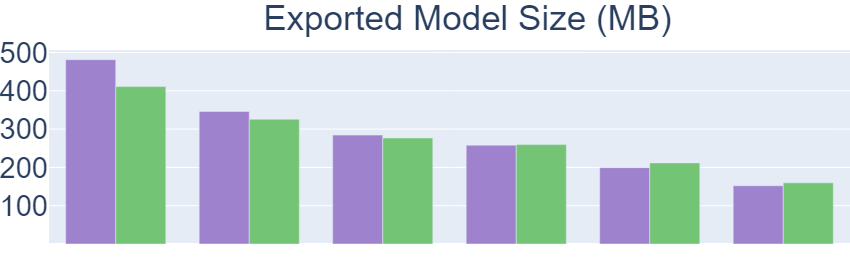}
	}\label{fig9a}
	\subfigure[MRI images.]{
    	\includegraphics[width=\linewidth]{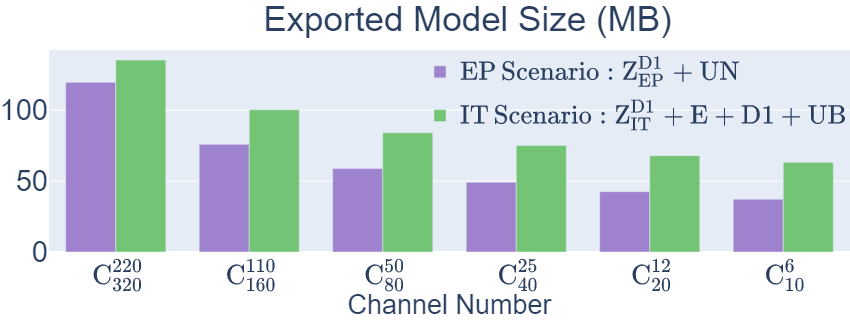}
	}\label{fig9b}
\caption{The size of the exported network when stealing 100 images with the dictionary approach. In the EP Scenario, the exported network includes the compression codes $Z$ and the utility network UN, while in the IT Scenario, it includes the compression codes $Z$, the decoder D1, the encoder E and utility branch UB from the data lake.
}
\label{fig9}
\end{figure}

In the dictionary method, the compression codes are stored as a key-value pairs within the checkpoint of the exported network. This does not alter the network parameters, but increases the size of the exported network. In Fig.~\ref{fig9}, the size of the exported model when exfiltrating 100 CT or MR images are displayed  as a function of the number of latent channels and scenario type. In the EP Scenario, one needs to export the codes and utility network whereas in the IT scenario one has to also export the encoder, the reduced decoder $D1$. The exported network is always below 500 MB, and even less than 140 MB for MR images.
Given that commonly used backbone models often exceed 500 MB in disk size (e.g., VGG16 is 576 MB), the size of the exported model is unlikely to raise suspicion from the data owner.

\subsection{Differential Privacy Mitigation}\label{sec4.5}
We study below the impact of differential privacy on the DEC attack with and without  decoded latent variables. We consider   a basic DP protocol that adds calibrated Gaussian noise with a zero mean and specified standard deviation to the exported model. 

\begin{figure}[htbp]
\centering
    \subfigure[CT images.]{
		\includegraphics[width=\linewidth]{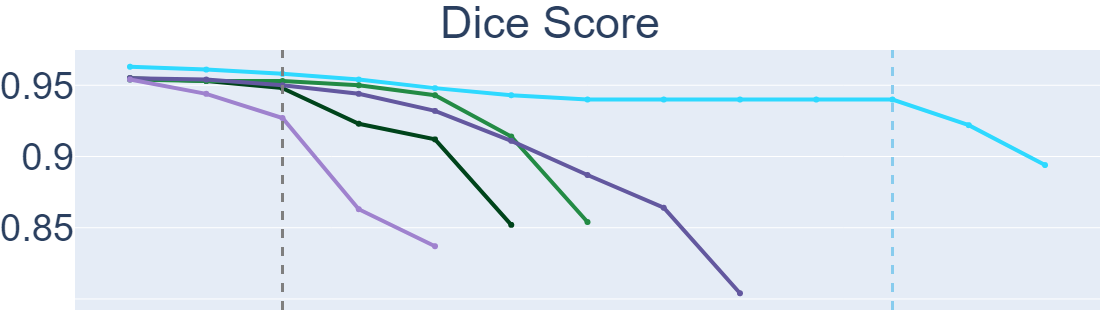}
	}\label{fig10a}
	\subfigure[MRI images.]{
    	\includegraphics[width=\linewidth]{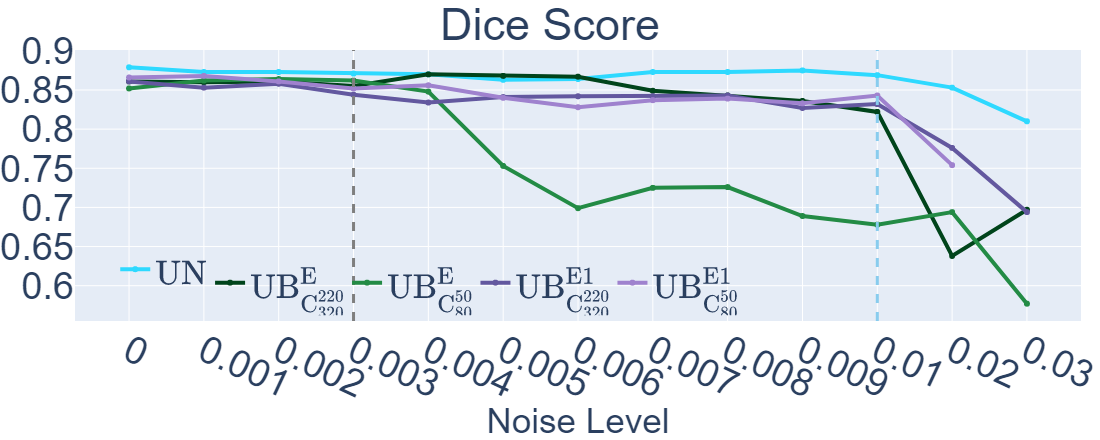}
	}\label{fig10b}
\caption{Utility task performances on CT and MRI images with varying noise levels applied to the utility model (UN for EP Scenario and UB for IT Scenario).  For the UB model, several encoders and channel numbers are considered.}
\label{fig10}
\end{figure}

\textbf{Noise Level Calibration.} It is crucial to carefully calibrate the noise standard deviation, as model performance is expected to decline with increasing noise levels. 
In Fig.~\ref{fig10}, we visualize the mean Dice score of the utility network (UN) or utility branch (UB) as a function of the noise level (i.e. standard deviation for Gaussian noise), the network architecture and the type of image to encode. We observe that a noise level less than  0.01 preserves the model performance in EP Scenario whereas a level of 0.002 (resp. 0.003) is a limit for the UB model of the IT Scenario trained on CT (resp. MR) images. The difference of noise sensitivity between the two scenarios could be explained by the fact that the encoder is not jointly trained with the UB in the IT scenario.

\textbf{Resilience of the DEC Attack to DP.}
As detailed in the Section~\ref{DPmitigation}, the attacker, knowing that the data owner applies  DP, is likely to adopt a different strategy by exporting the decoded latent variables instead of the encoded latent and hyperlatent variables, at the expense of stealing less images. 
We have performed several experiments to assess the resilience of the attack both in terms of  quality and quantity of the stolen images.

In the EP Scenario, the performance of the attack model seems unaffected by the application of  noise levels less than 1\%. We have consistently obtained a PSNR of 39.0 (resp. 36.9) and MS\_SSIM of 0.996 (resp. 0.997) for CT (resp. MR) images.
This resilience can be attributed to the inherent nature of the GAN-based HiFiC model where the decoded latent codes follow a Gaussian distribution  with a  standard deviation larger than 1\%.

\begin{figure}[htbp]
\centering
	\subfigure[CT images.]{
		\includegraphics[width=\linewidth]{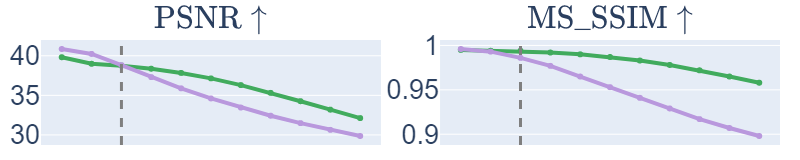}
	}\label{fig11a}
	\subfigure[MRI images.]{
    	\includegraphics[width=\linewidth]{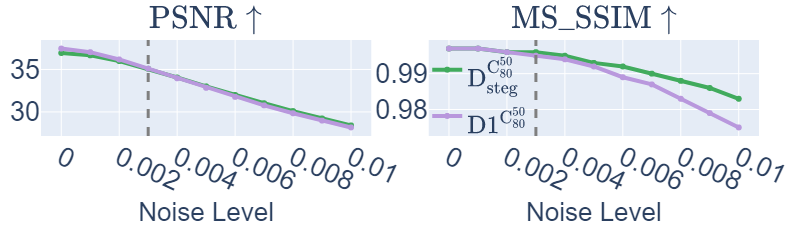}
	}\label{fig11b}
\caption{Lossy compression-based attack on CT and MRI images before and after applying Differential Privacy in IT Scenario, with varying noise levels applied to the attack decoder and the compression codes.}
\label{fig11}
\end{figure}

In IT Scenario, the noise addition perturbs both the compression codes and the decoder. From Fig.~\ref{fig11}, we see that the image quality of the stolen images remains acceptable when the noise level is below 0.003 (PSNR$>$35 and MS\_SSIM$>$0.98) for both image modalities. Besides, the architecture with the reduced decoder $D1$ is more sensitive to noise than the original full-size one. This is due to the dilution of noise created by the large ResNet blocks. 

It should be noted that this threshold of 0.003 is roughly the same as the noise level acceptable for the utility branch as displayed in Fig.~\ref{fig10}. This implies that the data owner is likely to apply a noise level of 0.003 or lower in order to preserve the performance of the exported network. But with this level of noise, the data owner will also preserve the quality of the exfiltrated images.

\begin{figure}[htbp]
    \centering
	\subfigure{
	\rotatebox{90}{\scriptsize{~~~~~~~CT}}
		\begin{minipage}[t]{0.17\linewidth}
			\centering
			\includegraphics[width=1\linewidth]{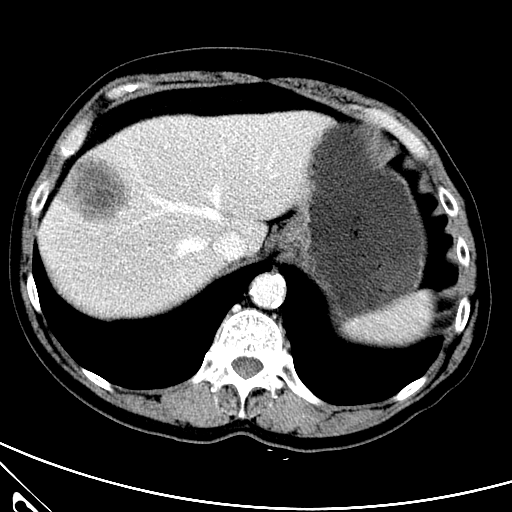}
		\end{minipage}
	}\hspace{-0.5\baselineskip}\vspace{-0.5\baselineskip}
	\subfigure{
		\begin{minipage}[t]{0.17\linewidth}
			\centering
			\includegraphics[width=1\linewidth]{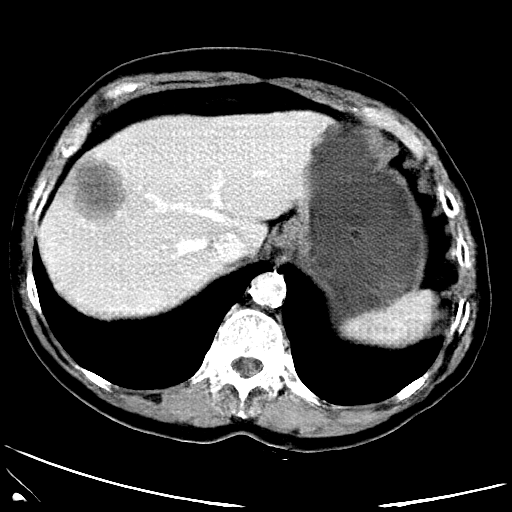}
		\end{minipage}
		\begin{minipage}[t]{0.17\linewidth}
			\centering
			\includegraphics[width=1\linewidth]{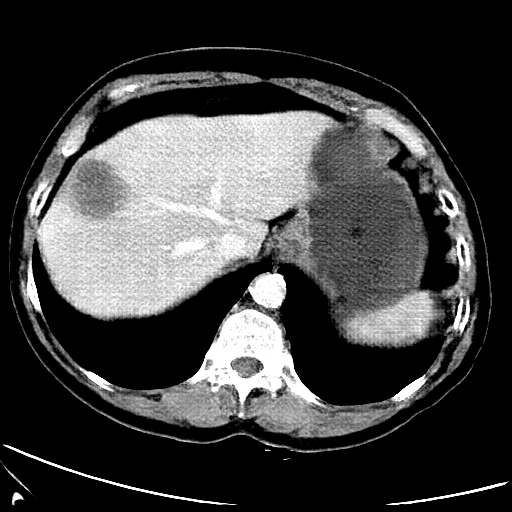}
		\end{minipage}
	}\hspace{-0.5\baselineskip}
	\subfigure{
		\begin{minipage}[t]{0.17\linewidth}
			\centering
			\includegraphics[width=1\linewidth]{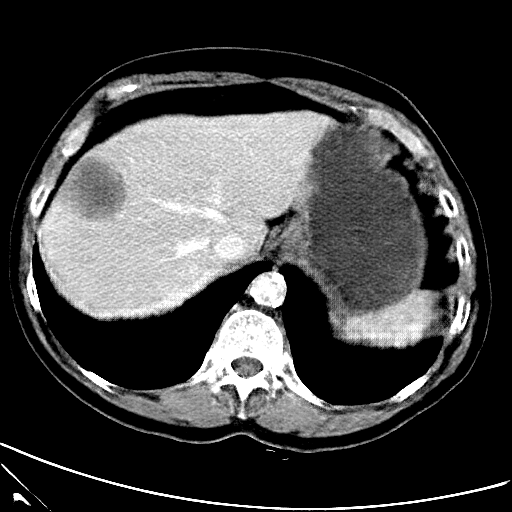}
		\end{minipage}
		\begin{minipage}[t]{0.17\linewidth}
			\centering
			\includegraphics[width=1\linewidth]{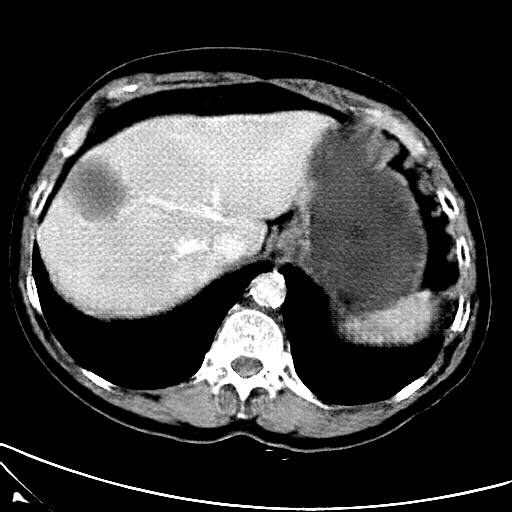}
			\end{minipage}
	}
	
	\subfigure{
		\begin{minipage}[t]{0.17\linewidth}
			\centering
			\includegraphics[width=1\linewidth]{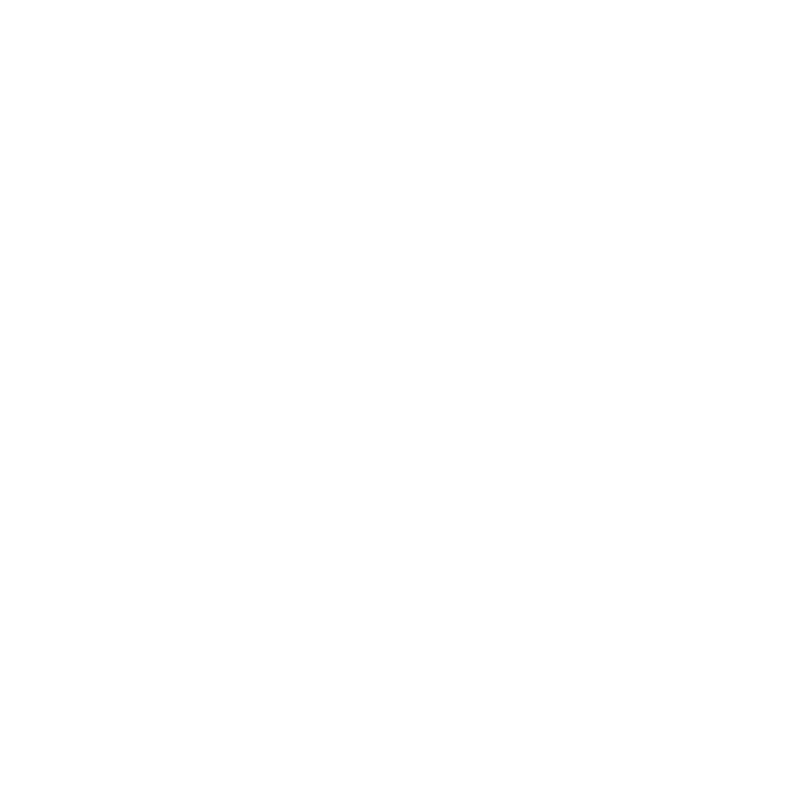}
		\end{minipage}
	}\hspace{-0.5\baselineskip}\vspace{-0.2\baselineskip}
	\subfigure{
	\rotatebox{90}{\scriptsize{~~~~CT DP}}
		\begin{minipage}[t]{0.17\linewidth}
			\centering
			\includegraphics[width=1\linewidth]{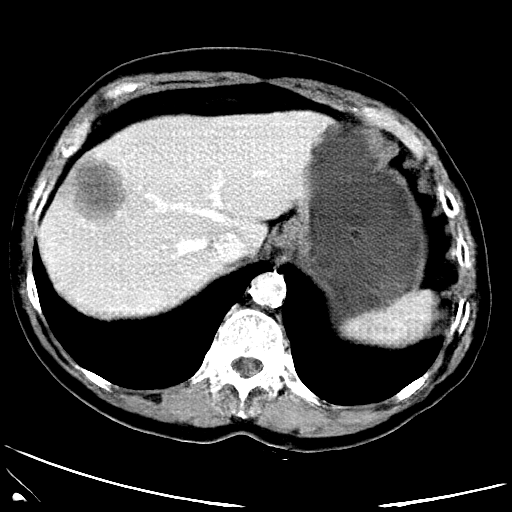}
		\end{minipage}
		\begin{minipage}[t]{0.17\linewidth}
			\centering
			\includegraphics[width=1\linewidth]{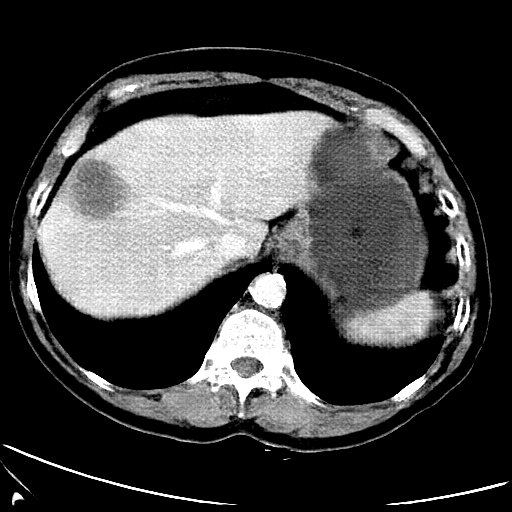}
		\end{minipage}
	}\hspace{-0.5\baselineskip}
	\subfigure{
		\begin{minipage}[t]{0.17\linewidth}
			\centering
			\includegraphics[width=1\linewidth]{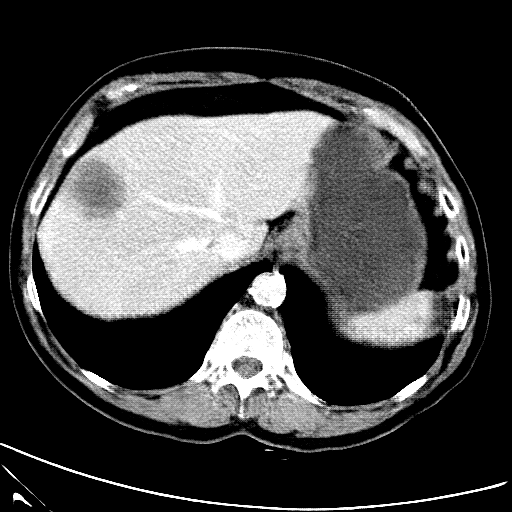}
		\end{minipage}
		\begin{minipage}[t]{0.17\linewidth}
			\centering
			\includegraphics[width=1\linewidth]{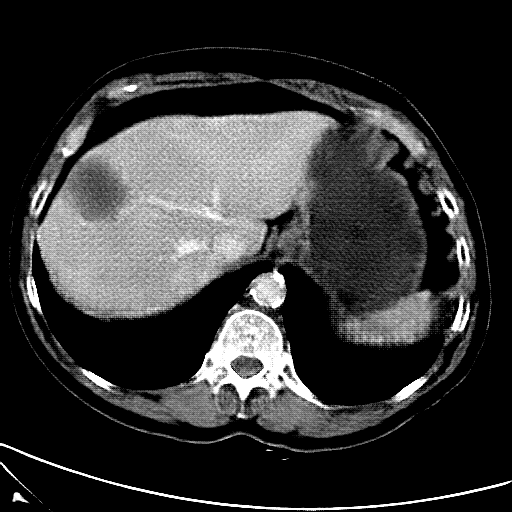}
		\end{minipage}
	}
	
	\subfigure{
	\rotatebox{90}{\scriptsize{~~~~~~MRI}}
		\begin{minipage}[t]{0.17\linewidth}
			\centering
			\includegraphics[width=1\linewidth]{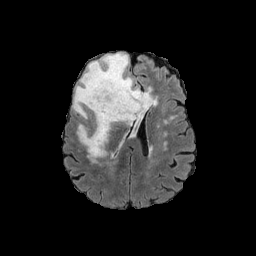}
		\end{minipage}
	}\hspace{-0.5\baselineskip}\vspace{-0.5\baselineskip}
	\subfigure{
		\begin{minipage}[t]{0.17\linewidth}
			\centering
			\includegraphics[width=1\linewidth]{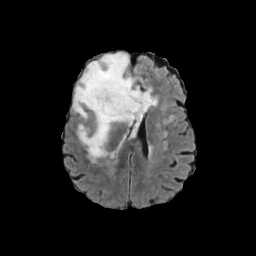}
		\end{minipage}
		\begin{minipage}[t]{0.17\linewidth}
			\centering
			\includegraphics[width=1\linewidth]{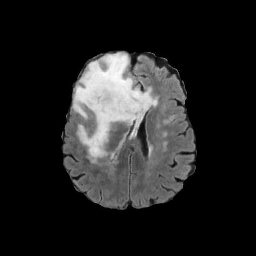}
		\end{minipage}
	}\hspace{-0.5\baselineskip}
	\subfigure{
		\begin{minipage}[t]{0.17\linewidth}
			\centering
			\includegraphics[width=1\linewidth]{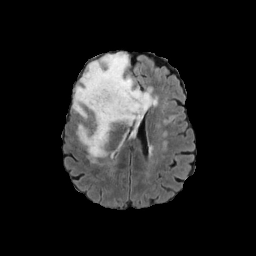}
		\end{minipage}
		\begin{minipage}[t]{0.17\linewidth}
			\centering
			\includegraphics[width=1\linewidth]{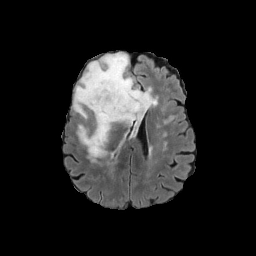}
		\end{minipage}
	}
	
	\subfigure{
		\begin{minipage}[t]{0.17\linewidth}
			\centering
			\includegraphics[width=1\linewidth]{./Figure/img/blank.png}
			\vspace{0.1\linewidth}\footnotesize$\Target$
		\end{minipage}
	}\hspace{-0.5\baselineskip}
    \setcounter{subfigure}{0}
	\subfigure[EP Scenario]{
	\rotatebox{90}{\scriptsize{~~~MRI DP}}
		\begin{minipage}[t]{0.17\linewidth}
			\centering
			\includegraphics[width=1\linewidth]{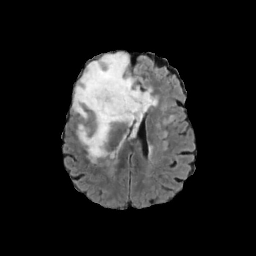}
			\vspace{0.1\linewidth}\footnotesize ${\Stolen}^{\mathrm{D}}$
		\end{minipage}
		\begin{minipage}[t]{0.17\linewidth}
			\centering
			\includegraphics[width=1\linewidth]{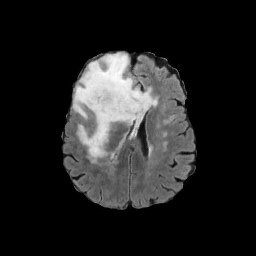}
			\vspace{0.1\linewidth}\footnotesize ${\Stolen}^{\mathrm{D1}}$
		\end{minipage}
	}\hspace{-0.5\baselineskip}\label{fig12a}
	\subfigure[IT Scenario]{
		\begin{minipage}[t]{0.17\linewidth}
			\centering
			\includegraphics[width=1\linewidth]{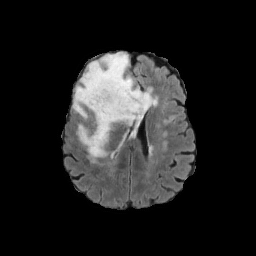}
			\vspace{0.1\linewidth}\footnotesize ${\Stolen}^{\mathrm{D}}$
		\end{minipage}
		\begin{minipage}[t]{0.17\linewidth}
			\centering
			\includegraphics[width=1\linewidth]{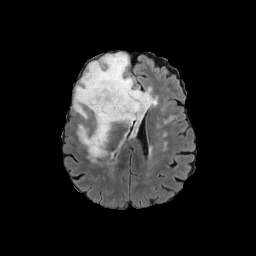}
			\vspace{0.1\linewidth}\footnotesize ${\Stolen}^{\mathrm{D1}}$
		\end{minipage}
	}\label{fig12b}
	\caption{Lossy image reconstructions of CT and MRI images before (row 1, 3) and after (row 2, 4) applying DP with calibrated noise levels. The first column displays the target images. The middle two columns show the results for the EP Scenario, while the last two columns is for the IT Scenario. For each scenario, we show the stolen images reconstructed by the decoder D or D1, respectively, utilizing a specific channel numbers ($\mathrm{C^{50}_{80}}$).
	}
	\label{fig12}
\end{figure}

The visual comparison of exfiltrated images before and after applying DP is displayed in Fig.~\ref{fig12}. 
In the EP Scenario, no noticeable differences are observed but in the IT Scenario, we observe slight variations of image intensity in the stolen CT images generated by the decoder D1. For MR images, no discernible differences are present. 

\begin{figure}[htbp]
\centering
	\subfigure[CT images.]{
		\includegraphics[width=\linewidth]{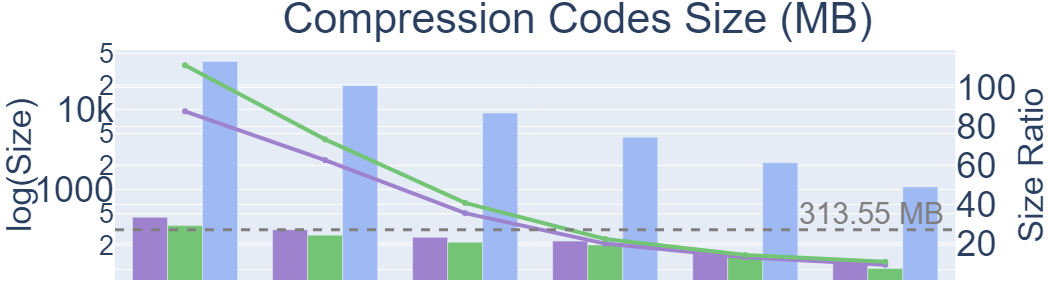}
	}\label{fig13a}
	\subfigure[MRI images.]{
    	\includegraphics[width=\linewidth]{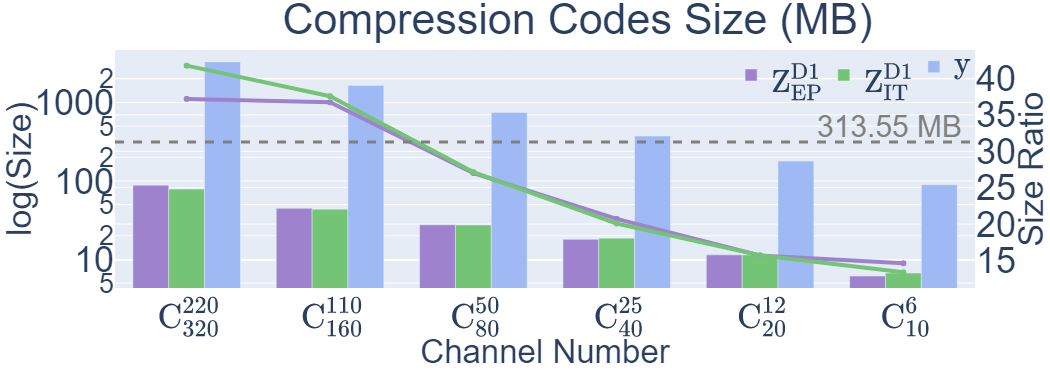}
	}\label{fig13b}
\caption{The size of the encoded and decoded compression codes. The bar chart with the left y-axis represents their sizes when stealing 100 images, while the line chart with the right y-axis illustrates the size ratio between $y$ and $Z$.
}
\label{fig13}
\end{figure}

\textbf{Size of the Exported Network.} 
Exporting decoded latent variables $y$ instead of encoded variables $Z$ results in an increase in the size of the compression codes.
This phenomenon is visible in Fig.~\ref{fig13}, where we show the size ratio $|y|/|Z|$ is around 40 for CT images and 30 for MRI images for the specific channel numbers ($\mathrm{C^{50}_{80}}$).
Specifically, the number of exfiltrated images is reduced by 40-fold for CT images and 30-fold for MRI images to ensure resilience against differential privacy measures. 
In a realistic scenario, if the exported model has disk size around 626 MB (using the full-size decoder $D$), then we can exfiltrate 313.55 MB  of  compression codes  (displayed by the gray dashed line in Fig.~\ref{fig13}), thus corresponding to the exfiltration of 
only 4 CT images or 42 MRI images after resorting to steganography.

\section{Discussion}\label{sec4.6}
\textbf{Discussion.} We have demonstrated that the DEC attack is realistic in the context of a medical data lake where the remote user is given access to CT or MR medical images with a utility task consisting in solving an image segmentation task. In both scenarios, an attacker can exfiltrate a large set of medical images when exporting a network solving a utility task (such as image detection, classification, registration, segmentation) outside the data lake. The attack feasibility is mainly a consequence of the two following facts:  i) neural networks for image analysis tend to be large in disk size, for instance with backbone models being larger than  500MB, and ii)  current deep compression networks allow one to reach an excellent fidelity / compression compromise.  We discuss below several key topics about the DEC attack.

\textbf{IT vs EP Scenarios.} The external pre-training scenario where a data encoder is imported inside the data lake is far simpler to handle for an attacker than the internal training scenario. Indeed, in this case, the attacker has only to produce the compression codes and hide them inside the exported network. The risk of detection is mainly related to the import of the encoder that must be hidden from the data owner. In the alternative IT scenario, the risk of detection occurs during the export of the model due to the dual heads of the model. Both risks are related to the ability of the attacker to hide the real nature of the model.

\textbf{Nature of Utility Network.} While previously proposed attacks such as the model inversion~\cite{10003239} or transpose~\cite{amit2024transposeattackstealingdatasets} attacks were assuming that the exported model was a classifier (without skip connections), the DEC attack does not put any constraint on the architecture of the utility model or the utility branch. In the EP scenario, the network is only a recipient to hiding compression codes, while in the IT scenario, we have shown that the data encoder can serve as a backbone from which a task specific head (solving image segmentation in our case) can be connected. This greatly expands the applicability of the attack compared to previously proposed ones.

\textbf{Effective Compression Network.}  Based on the HiFiC architecture, we have optimized the performance / size ratio of this deep compression network based on a reduced decoder $D1$, and a limited number of channels of the latent and hyperlatent variables. With this configuration, our compression model was able to produce  CT and MR images of high  quality with compressed files that are around 30-50 times smaller than the size obtained with lossless compression (e.g. gzip algorithm) with a decoder weighting less than 30MB. The research on universal deep image compression\cite{tsubota2023universal} is very active, and it is expected that the trade-off between fidelity and compression will continue to improve favorably with higher quality images from smaller codes. For instance, one way to improve the HiFiC network would be to design a more efficient entropy coding~\cite{hu2021learning} module, for instance by incorporating a deep prior about the distribution of the latent variables $\zlatent$.

\textbf{Resilience to Differential Privacy.} Given the considered scenarios, we have studied the addition of noise on the model parameters rather than on the images or the gradient during training. The addition of Gaussian noise to the exported model has a drastic effect on the DEC attack, making it impossible to reconstruct any image, if the hidden compression codes $\zlatent$ are the encoded latent variables. However, when exporting the decoded latent $y$ instead of  $\zlatent$, the attacker can reconstruct a limited number of stolen images with sufficient quality for a certain range of noise level. Besides, we found that the amount of noise necessary to prevent image reconstruction is higher than the noise level seriously decreasing the performance of the utility model. This leads the data owner to a difficult dilemma: either exporting a secured but seriously limited model, or exporting an effective model but with a risk of data exfiltration. Obviously, the compromise made by the data owner depends on the trustworthiness of the data user.

\begin{figure}[!htbp]
\centering
\includegraphics[width=\linewidth]{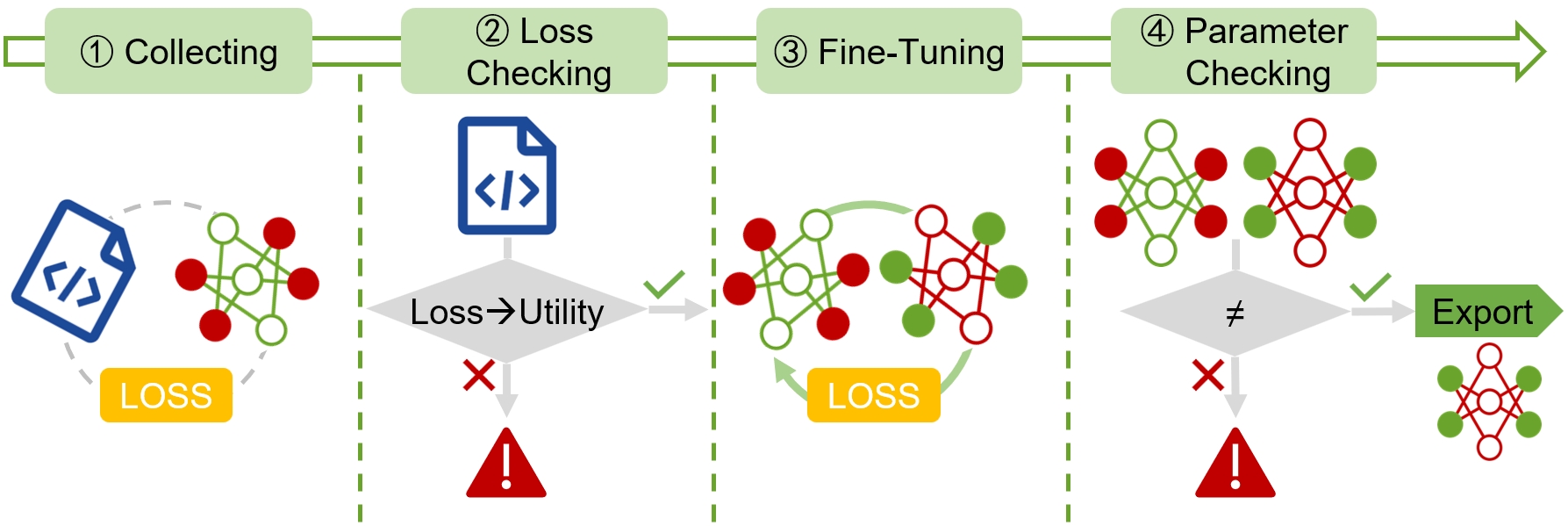}
\caption{Flowchart for preventing data exfiltration through fine-tuning. The data owner implements this prevention strategy within the data lake in four steps. 
\ding{172}Collecting: Gather the source code for the model and the loss functions required for fine-tuning. 
\ding{173}Loss Checking: Review the source code and confirm whether the loss functions align with the utility task objectives. If they do, the process proceeds to the next step; otherwise, an alert is triggered.
\ding{174}Fine-Tuning: Refine the model until all parameters have significantly deviated from their initial values. 
\ding{175}Parameter Checking: Check if all parameters have changed after fine-tuning. If any remain unchanged, an alert is raised. If all parameters are modified, the fine-tuned model is approved for export from the data lake.}
\label{fig14}
\end{figure}

\textbf{Prevention based on Model Fine-tuning.}  To mitigate the risk of data leakage when exporting a machine learning model from a data lake, we propose an alternative to differential privacy based on model fine-tuning. The prevention process, as illustrated in Fig.~\ref{fig14}, involves four key steps executed by the data owner within the data lake. First, the data owner collects the source code for the model intended for export and the associated loss functions required for fine-tuning on the training data. Second, the source code is reviewed to ensure that the loss functions align with the utility task objectives. If alignment is confirmed, the model is fine-tuned  until all model parameters have  been sufficiently modified. If some parameters have not been modified after a fixed number of epochs, then an alert is raised. If all stages have been cleared, the model is exported.  Otherwise, the data owner does not proceed with  the model export and start discussing with the data user about potential security issues in the model. 

We believe that this export process enforces privacy with existing data exfiltration attacks while preserving the utility of the model. Indeed, similarly to differential privacy, the proposed process modifies the model parameters thus making it impossible to hide compression codes with it. Besides it allows to detect secondary branches in the utility model that are not modified during the fine-tuning but that can serve to decode compression codes. Finally, by checking the nature of the loss functions optimized during fine-tuning, the data owner can check the relevance  of those objective functions with respect to the utility task and at this occasion, can detect most attacks including the transpose attack~\cite{amit2024transposeattackstealingdatasets}. The advantage of this Fine-Tuning prevention is that  data privacy is not obtained at the expense of the model performance, as limited fine-tuning over a few epochs should not substantially alter its effectiveness. Additionally, it is not intrusive for the data user, but it does require the data owner to have the capability to understand the code related to the loss functions, fine-tune the model, and monitor any modifications to the model parameters.   

\section{Conclusion}
In this paper, we have proposed a  data exfiltration attack  on medical images  consisting in using pretrained or learned compression / decompression algorithms. We have shown that this is a  realistic and effective attack if a remote user acts as an attacker. We have demonstrated that by combining different methods (training smaller model from scratch, steganography, latent channel downsizing...) it is possible to decrease the model size, and the compression codes while preserving the quality of stolen images and exporting an effective utility model. For instance, with learned lossy image compression, it is possible to obtain compression codes that are 60 times smaller than the ones obtained with baseline lossless compression. 
The number of images that can be stolen is on the order of at least one hundred images for an exported model size of 300 MB. Importing a  pretrained model in the data lake does not have a large influence on the attack performance compared to the internal training scenario. The quality of the stolen images with this attack is high (MS\_SSIM around 0.996) and can be used to successfully train an utility model outside a data lake. 
Additionally, we have evaluated several mitigation methods against the proposed DEC attack in a recent publication~\cite{thellier:hal-05167639}, as well as against another data exfiltration technique, namely the transpose attack~\cite{amit2024transposeattackstealingdatasets}. The experiments were conducted on three medical datasets. Results indicate that while the transpose attack can be successfully applied to 2D medical images for classification tasks, it struggles to generate high-quality images when the resolution increases to 512$\times$512, as in the case of the MIMIC-CXR dataset.

To prevent this attack,  the data owner should use multi-factor authentication combined with some monitoring of the user activity inside the data lake, although it may be difficult to implement efficiently. The use of different privacy is effective when encoded latent and hyperlatent codes are exported within the model. But, we have shown that the attacker can make their attack resilient to a simple differential privacy protection by  exporting decoded latent variables instead of encoded latent and hyperlatent variables. Finally, we have introduced an export process that can be implemented by a data owner to prevent the data exfiltration by compression attack. This process is based on the fine-tuning of a model until all model parameters have significantly changed, and the  inspection of the source codes providing the loss functions. A comprehensive evaluation of the effectiveness of this fine-tuning-based mitigation approach is presented in our recent publication~\cite{thellier:hal-05167639}. 

Future work will explore a variety of privacy-preserving techniques, including knowledge distillation~\cite{li2022swing}, homomorphic encryption, and the generation of anonymized imaging data from original ones.


\acks{This work has been supported by the French government, through
the National Research Agency (ANR) 3IA Côte d’Azur and IA Cluster project (ANR-19-3IA-0002 and ANR-23-IACL-0001). The authors are grateful to the OPAL infrastructure from Université Côte d’Azur for providing resources and support.

We would like to thank Teddy Furon for his invaluable suggestions on steganography, as well as his insightful comments.}

%
\ethics{The work follows appropriate ethical standards in conducting research and writing the manuscript, following all applicable laws and regulations regarding treatment of animals or human subjects.}

\coi{We declare we don't have conflicts of interest.}

\data{The data is publicly available.}

\bibliography{sample}


\end{document}